\documentclass[aps,prc,reprint,superscriptaddress]{revtex4-1}
\usepackage{dcolumn}
\usepackage{amsmath,amssymb,amsthm,paralist}
\usepackage{graphicx}
\usepackage{appendix}
\usepackage{subfigure}
\usepackage{longtable}
\usepackage{url}
\usepackage{verbatim}

\begin{document}


\title{Time-of-flight mass measurements of neutron-rich chromium
isotopes up to $N=40$ and implications for the accreted neutron star crust}


\author{Z.~Meisel}
\email[]{zmeisel@nd.edu}
\affiliation{Department of Physics, University of Notre Dame, Notre Dame, Indiana 46556, USA}
\affiliation{Joint Institute for Nuclear Astrophysics, Michigan State University, East Lansing, Michigan 48824, USA}
\author{S.~George}
\affiliation{Joint Institute for Nuclear Astrophysics, Michigan State University, East Lansing, Michigan 48824, USA}
\affiliation{Max-Planck-Institut f\"{u}r Kernphysik, Heidelberg 69117, Germany}
\author{S.~Ahn}
\affiliation{Joint Institute for Nuclear Astrophysics, Michigan State University, East Lansing, Michigan 48824, USA}
\affiliation{National Superconducting Cyclotron Laboratory, Michigan State University, East Lansing, Michigan 48824, USA}
\author{D.~Bazin}
\affiliation{National Superconducting Cyclotron Laboratory, Michigan State University, East Lansing, Michigan 48824, USA}
\author{B.A.~Brown}
\affiliation{National Superconducting Cyclotron Laboratory, Michigan State University, East Lansing, Michigan 48824, USA}
\affiliation{Department of Physics and Astronomy, Michigan State University, East Lansing, Michigan 48824, USA}
\author{J.~Browne}
\affiliation{Joint Institute for Nuclear Astrophysics, Michigan State University, East Lansing, Michigan 48824, USA}
\affiliation{National Superconducting Cyclotron Laboratory, Michigan State University, East Lansing, Michigan 48824, USA}
\affiliation{Department of Physics and Astronomy, Michigan State University, East Lansing, Michigan 48824, USA}
\author{J.F.~Carpino}
\affiliation{Department of Physics, Western Michigan University, Kalamazoo, Michigan 49008, USA}
\author{H.~Chung}
\affiliation{Department of Physics, Western Michigan University, Kalamazoo, Michigan 49008, USA}
\author{R.H.~Cyburt}
\affiliation{Joint Institute for Nuclear Astrophysics, Michigan State University, East Lansing, Michigan 48824, USA}
\affiliation{National Superconducting Cyclotron Laboratory, Michigan State University, East Lansing, Michigan 48824, USA}
\author{A.~Estrad\'{e}}
\affiliation{Joint Institute for Nuclear Astrophysics, Michigan State University, East Lansing, Michigan 48824, USA}
\affiliation{Department of Physics, Central Michigan University, Mount Pleasant, Michigan 48859, USA}
\author{M.~Famiano}
\affiliation{Department of Physics, Western Michigan University, Kalamazoo, Michigan 49008, USA}
\author{A.~Gade}
\affiliation{National Superconducting Cyclotron Laboratory, Michigan State University, East Lansing, Michigan 48824, USA}
\affiliation{Department of Physics and Astronomy, Michigan State University, East Lansing, Michigan 48824, USA}
\author{C.~Langer}
\affiliation{Institute for Applied Physics, Goethe University
Frankfurt a. M., Frankfurt a. M. 60438, Germany}
\author{M.~Mato\v{s}}
\affiliation{Physics Section, International Atomic Energy Agency, Vienna 1400, Austria}
\author{W.~Mittig}
\affiliation{National Superconducting Cyclotron Laboratory, Michigan State University, East Lansing, Michigan 48824, USA}
\affiliation{Department of Physics and Astronomy, Michigan State University, East Lansing, Michigan 48824, USA}
\author{F.~Montes}
\affiliation{Joint Institute for Nuclear Astrophysics, Michigan State University, East Lansing, Michigan 48824, USA}
\affiliation{National Superconducting Cyclotron Laboratory, Michigan State University, East Lansing, Michigan 48824, USA}
\author{D.J.~Morrissey}
\affiliation{National Superconducting Cyclotron Laboratory, Michigan State University, East Lansing, Michigan 48824, USA}
\affiliation{Department of Chemistry, Michigan State University, East Lansing, Michigan 48824, USA}
\author{J.~Pereira}
\affiliation{Joint Institute for Nuclear Astrophysics, Michigan State University, East Lansing, Michigan 48824, USA}
\affiliation{National Superconducting Cyclotron Laboratory, Michigan State University, East Lansing, Michigan 48824, USA}
\author{H.~Schatz}
\affiliation{Joint Institute for Nuclear Astrophysics, Michigan State University, East Lansing, Michigan 48824, USA}
\affiliation{National Superconducting Cyclotron Laboratory, Michigan State University, East Lansing, Michigan 48824, USA}
\affiliation{Department of Physics and Astronomy, Michigan State University, East Lansing, Michigan 48824, USA}
\author{J.~Schatz}
\affiliation{National Superconducting Cyclotron Laboratory, Michigan State University, East Lansing, Michigan 48824, USA}
\author{M.~Scott}
\affiliation{National Superconducting Cyclotron Laboratory, Michigan State University, East Lansing, Michigan 48824, USA}
\affiliation{Department of Physics and Astronomy, Michigan State University, East Lansing, Michigan 48824, USA}
\author{D.~Shapira}
\affiliation{Oak Ridge National Laboratory, Oak Ridge, Tennessee 37831, USA}
\author{K.~Sieja}
\affiliation{Universit\'e de Strasbourg, IPHC, CNRS, UMR7178, Strasbourg 67037, France}
\author{K.~Smith}
\affiliation{Department of Physics and Astronomy, University of Tennessee, Knoxville, Tennessee 37996, USA}
\author{J.~Stevens}
\affiliation{Joint Institute for Nuclear Astrophysics, Michigan State University, East Lansing, Michigan 48824, USA}
\affiliation{National Superconducting Cyclotron Laboratory, Michigan
State University, East Lansing, Michigan 48824, USA}
\affiliation{Department of Physics and Astronomy, Michigan State University, East Lansing, Michigan 48824, USA}
\author{W.~Tan}
\affiliation{Department of Physics, University of Notre Dame, Notre Dame, Indiana 46556, USA}
\author{O.~Tarasov}
\affiliation{National Superconducting Cyclotron Laboratory, Michigan State University, East Lansing, Michigan 48824, USA}
\author{S.~Towers}
\affiliation{Department of Physics, Western Michigan University, Kalamazoo, Michigan 49008, USA}
\author{K.~Wimmer}
\affiliation{Department of Physics, University of Tokyo, Hongo 7-3-1, Bunkyo-ku, Tokyo 113-0033, Japan}
\author{J.R.~Winkelbauer}
\affiliation{National Superconducting Cyclotron Laboratory, Michigan State University, East Lansing, Michigan 48824, USA}
\affiliation{Department of Physics and Astronomy, Michigan State University, East Lansing, Michigan 48824, USA}
\author{J.~Yurkon}
\affiliation{National Superconducting Cyclotron Laboratory, Michigan State University, East Lansing, Michigan 48824, USA}
\author{R.G.T.~Zegers}
\affiliation{Joint Institute for Nuclear Astrophysics, Michigan State University, East Lansing, Michigan 48824, USA}
\affiliation{National Superconducting Cyclotron Laboratory, Michigan State University, East Lansing, Michigan 48824, USA}
\affiliation{Department of Physics and Astronomy, Michigan State University, East Lansing, Michigan 48824, USA}


\date{\today}

\begin{abstract}
We present the mass excesses of $^{59-64}$Cr, obtained from recent time-of-flight nuclear mass
measurements at the National Superconducting Cyclotron Laboratory at
Michigan State University. The mass of $^{64}$Cr was determined for
the first time with an atomic mass excess of  $-33.48(44)$~MeV. We find a significantly different
two-neutron separation energy $S_{2n}$ trend for neutron-rich
isotopes of chromium, removing the previously observed enhancement in
binding at $N=38$. Additionally, we extend the $S_{2n}$ trend for
chromium to $N=40$, revealing behavior consistent with the
previously identified island of inversion in this region. We compare
our results to state-of-the-art shell-model calculations performed
with a modified Lenzi-Nowacki-Poves-Sieja interaction in the
$fp$-shell, including the $g_{9/2}$ and $d_{5/2}$ orbits for the
neutron valence space.
We employ our result for the mass of $^{64}$Cr in 
accreted neutron star crust network calculations and find a
reduction in the strength and depth of electron capture
heating from the $A=64$ isobaric chain, resulting in a cooler than
expected accreted neutron star crust.
This reduced heating is found to be 
due to the over 1~MeV reduction in binding for $^{64}$Cr with
respect to values from commonly
used global mass models.
\end{abstract}


\maketitle

\section{Introduction}
\label{sec:Intro}
The evolution of nuclear structure away from the valley of
$\beta$-stability is a direct consequence of the forces at work in
nuclei~\cite{Brow01,Jans13}. Neutron-rich nuclides are of particular
interest, since much of the neutron-rich nuclear landscape has
yet to be explored~\cite{Erle12}. Recently, the experimental reach of
radioactive ion beam facilities has extended to chromium for neutron
number $N=40$, where an island of inversion has been inferred
from various experimental
signatures~\cite{Sorl03,Gade10,Baug12,Naim12,Craw13,Brau15}. Trends
in first $2^{+}$ excited state energies $E(2_{1}^{+})$ and ratios between
first $4^{+}$ excited state energies and $E(2_{1}^{+})$ demonstrated a
structural change between iron (proton number $Z=26)$ and chromium ($Z=24)$
isotopes near $N=40$~\cite{Sorl03,Adri08,Gade10,Lenz10}. This increase in
collectivity for chromium near $N=40$, attributed to a rapid shape
change from spherical to deformed structures, is further supported by
quadrupole excitation strength $B(E2)$
measurements~\cite{Baug12,Craw13,Brau15}.
Nuclear mass measurements provide an independent probe of
structural evolution which, in contrast to $B(E2)$ measurements, can
avoid the bias to proton degrees of freedom~\cite{Lunn03,Meis15}.
Precision mass measurements of manganese isotopes have indicated
that the $N=40$ sub-shell gap has broken down by
$Z=25$~\cite{Naim12}. However, mass measurements have yet to extend
to $N=40$ in the chromium isotopic chain.

The $N=40$ chromium isotope $^{64}$Cr is of astrophysical interest
due to the expected prevalence of $A=64$ material 
on the surfaces of accreting neutron
stars, and therefore in the outer neutron star
crust~\cite{Scha99,Scha01,Scha03}. The trend in nuclear masses along
an isobaric chain strongly impacts the depth and strength of
electron capture reactions that heat and cool the outer crust,
altering its thermal profile~\cite{Gupt07,Estr11,Meis15b}. The
resultant thermal profile impacts a host of astronomical
observables, including the ignition of type-I x-ray
bursts~\cite{Woos76,Scha06,Pari13}
and superbursts~\cite{Cumm01,Keek12}, cooling of transiently accreting neutron
stars while accretion is turned off~\cite{Brow09,Deib15}, and
potentially gravitational wave emission~\cite{Bild98,Usho00}.

To investigate the open questions in nuclear structure and
astrophysics regarding the neutron-rich chromium
isotopes, we performed time-of-flight (TOF) mass measurements of
$^{59-64}$Cr ($Z=24,N=35-40$) at the
National Superconducting Cyclotron Laboratory (NSCL) at Michigan State
University. Argon and scandium mass measurements that were a part of
the same experiment are discussed in
Refs.~\cite{Meis15}~and~\cite{Meis15b}, respectively.
These new chromium 
masses show significant deviations from the
chromium mass trend presented in the 2012 Atomic Mass
Evaluation~\cite{Audi12}, implying a different structural evolution
along the chromium isotopic chain. Our mass measurement of $^{64}$Cr
extends the mass trend of chromium out to $N=40$ for the first time.
We employ this $^{64}$Cr mass in accreted neutron star crust
reaction network calculations and, due to the reduction in binding
of $^{64}$Cr compared to global mass models, find less heating and
shallower heating depths than previously expected.
\begin{figure*}[ht]\begin{center}
\includegraphics[width=1.8\columnwidth,angle=0]{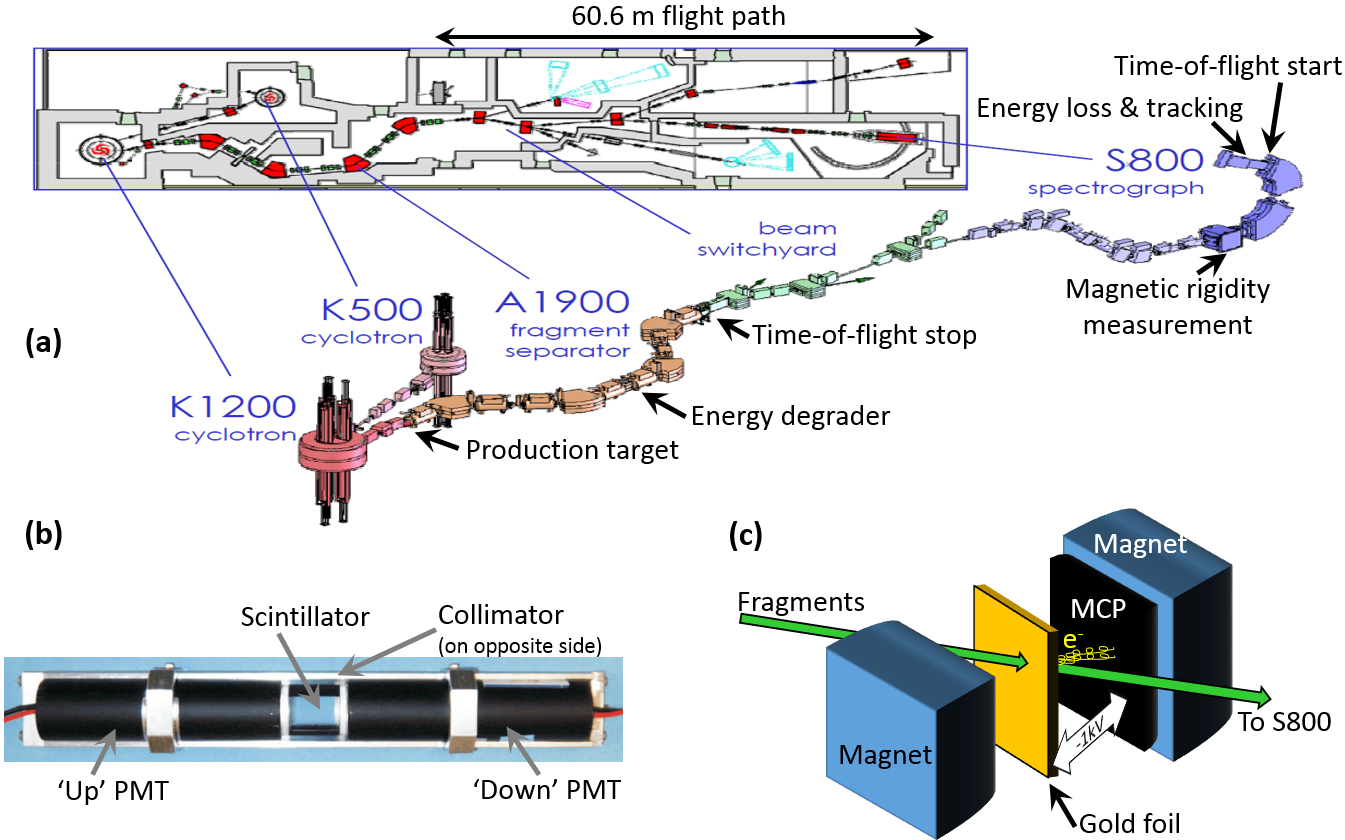}
\caption
{(color online.) (a) Schematic of the NSCL time-of-flight (TOF) mass
measurement set-up. (b) Scintillator and photomultiplier tube (PMT)
pair used to measure TOF stop and start signals at the A1900 and
S800 focal planes, respectively. Note that the delayed timing signal
from the A1900 was chosen as the stop signal to avoid triggering on
events which did not traverse the full flight path. (c) Schematic of the rigidity
measurement set-up at the target position of the S800. The
green arrow represents the beam fragments and the yellow spirals
represent the
secondary electrons fragments produce by passing through the gold
foil, which follow a helical trajectory towards the microchannel
plate detector (MCP) due to the -1~kV bias and magnetic field
established by the permanent magnets.
}
\label{fig:Setup}
\end{center}
\end{figure*}

\begin{figure}[ht]\begin{center}
\includegraphics[width=1.0\columnwidth,angle=0]{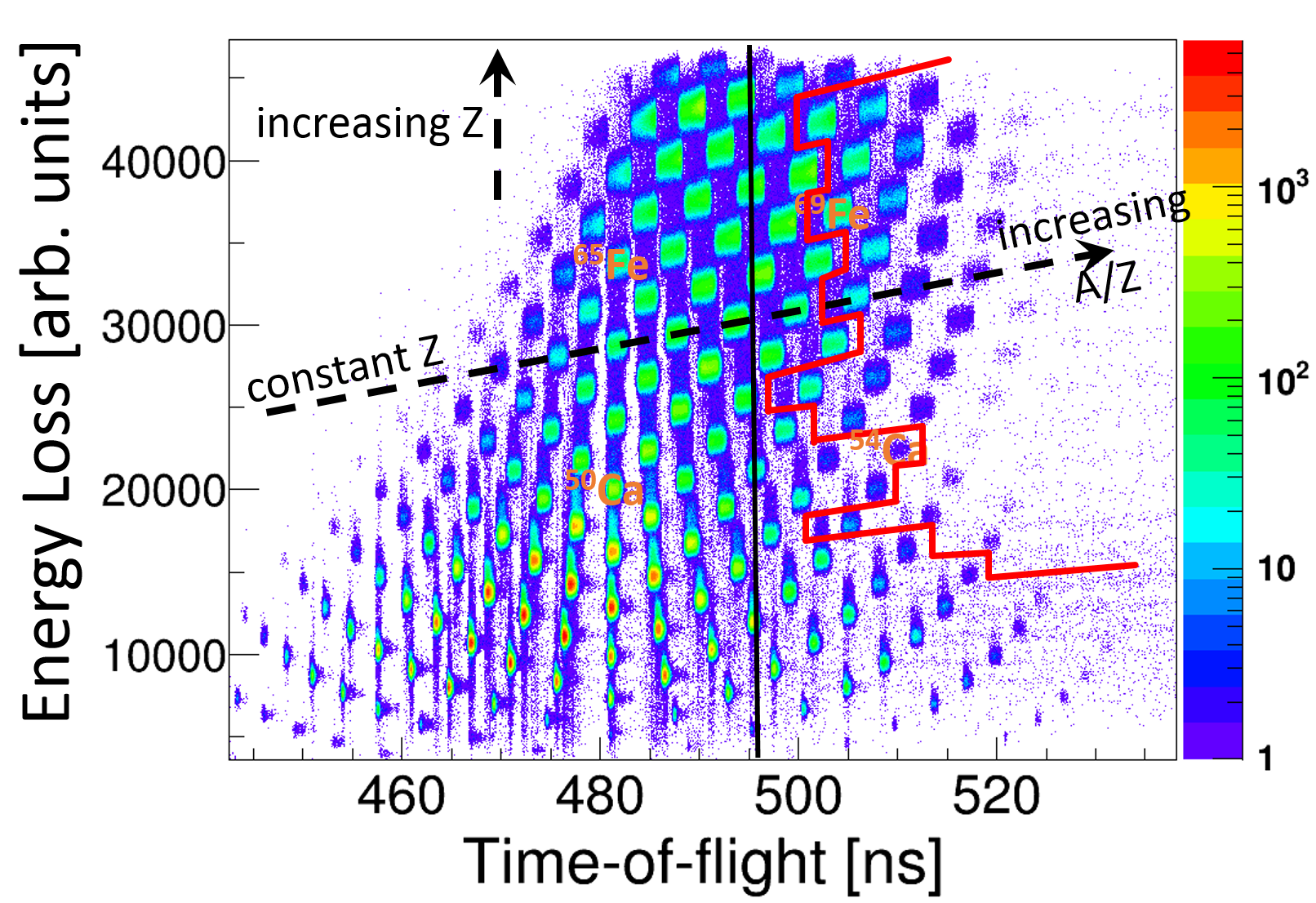}
\caption
{(color online.) Particle
identification plot of nuclei produced
in this time-of-flight (TOF)
mass-measurement experiment, where the color indicates production
intensity (counts per
100~picoseconds$\times$10~ionization-chamber-adc-units) and TOF
was not rigidity-corrected. Nuclei located to the right of the
red-line had no known experimental mass prior to the
mass-measurement reported here; $^{50}$Ca, $^{54}$Ca, $^{65}$Fe, and
$^{69}$Fe are labeled for reference.
 The
 data are from $\approx11$~hours of thin-target production and
 $\approx91$~hours of thick-target production.
}
\label{fig:LabeledPID}
\end{center}
\end{figure}

\section{Experimental Set-up}
\label{sec:ExptSetup}

\subsection{Time-of-flight mass measurement technique}
\label{ssec:TOFtechnique}
The masses presented in this work were measured via the
time-of-flight (TOF) technique, in which the flight times of ions
through a magnetic beam line system
are converted to nuclear masses by
comparison to the flight times of nuclides with known
masses~\cite{Meis13}. This technique was chosen due to its
ability to obtain masses for exotic nuclides at the
frontier of the known mass surface~\cite{Estr11,Gaud12}. We employed
the TOF mass measurement set-up at the NSCL at Michigan State
University described in detail in Ref.~\cite{Mato12}, shown in
Fig.~\ref{fig:Setup}.
This set-up consists of a 60.6~m flight path between the A1900
fragment separator~\cite{Morr03} and S800
spectrograph~\cite{Bazi03}, with fast-timing detectors located at
the A1900 and S800 focal planes, magnetic rigidity $B\rho$ detection at the
S800 target position, and energy loss and tracking detectors at the
S800 focal plane~\cite{Yurk99}. About 150 neutron-rich isotopes of
silicon to zinc were measured simultaneously
over the course of $\sim$100~hours.

The Coupled Cyclotron Facility~\cite{York99} at the NSCL was used to produce a
140~MeV/$u$ $^{82}$Se$^{32+}$ primary beam with an intensity of
$\sim$30 particle nA, which was fragmented on a
beryllium target to produce nuclei of interest. Target thicknesses
of 517~mg$\,$cm$^{-2}$, for production of
less neutron-rich calibration nuclides,
and 658~mg$\,$cm$^{-2}$, for production of more neutron-rich
nuclides of interest were used alternately, keeping $B\rho$ of the
A1900 and S800 fixed.
Fragments
were transmitted through the A1900 fragment separator~\cite{Bazi03},
where slits reduced the momentum acceptance to $\pm0.5$\%. A
7.2~mg$\,$cm$^{-2}$ Kapton wedge degrader was placed at the intermediate
image of the A1900 to remove the high-flux of low-$Z$ fragments that would have
otherwise
complicated particle identification (PID) and increased data
acquisition dead-time.
The S800 analysis line ion optics
were set to a dispersion-matching mode to provide a momentum
dispersion at the S800 target position of $\approx$1\%/11~cm that
enables an accurate rigidity measurement. This ion optical setting
provides an achromatic focus on the timing detectors in the A1900
and S800 focal planes. The
full set of nuclei detected over the course of the mass measurement
is shown in Fig.~\ref{fig:LabeledPID}. Timing and magnetic rigidity determinations will be discussed in more detail in Sections~\ref{ssec:Timing} and \ref{ssec:Rigidity}, respectively.

The relationship between TOF and nuclear rest mass $m_{\rm{rest}}$
is obtained from the equation of motion for a charged massive particle through a magnetic system. Equating the two counteracting forces, the Lorentz force $F_{\rm{L}}$ and the centripetal force $F_{\rm{c}}$, results in the following relationship:
\begin{eqnarray}
F_{\rm{c}}&=&F_{\rm{L}}\nonumber\\
\frac{\gamma(v) m_{\rm{rest}}v^{2}}{\rho}&=&qvB\nonumber\\
m_{\rm{rest}}&=&\frac{1}{v}\frac{q(B\rho)}{\gamma(v)}\nonumber\\
m_{\rm{rest}}&=&\frac{\rm{TOF}}{L_{\rm{path}}}\frac{q(B\rho)}{\gamma\left(\frac{L_{\rm{path}}}{\rm{TOF}}\right)},
\label{eqn:TOFmassRelationship}
\end{eqnarray}
where the Lorentz factor $\gamma$ is a function of velocity
$v$, which is in turn the ratio of flight-path length
$L_{\rm{path}}$ to flight time TOF. It follows that, in principle,
the simultaneous measurement of an ion's TOF, charge $q$, and
$B\rho$ through a system of known $L_{\rm{path}}$ yields
$m_{\rm{rest}}$. However, in practice $L_{\rm{path}}$ and the ion
optical dispersion used to measure $B\rho$ are not known with
sufficient precision to obtain a precise value for $m_{\rm{rest}}$.
Furthermore, it is more practical to make a relative than an
absolute measurement of
$B\rho$.
Instead, the
$\frac{m_{\rm{rest}}}{q}\left(\rm{TOF}\right)$ relationship is determined
empirically by measuring the TOF of calibration or reference
nuclides~\cite{Meis13}. The chosen reference nuclides have well-known masses
($\lesssim100$~keV uncertainty), no
known isomers with lifetimes comparable to the flight time
($\sim$500~ns), and are
as close as possible in
nuclear charge $Z$ and mass $A$ to the nuclides of interest in
order to minimize systematic uncertainties~\cite{Meis13}.

\begin{figure*}[]\begin{center}
\includegraphics[width=2.0\columnwidth,angle=0]{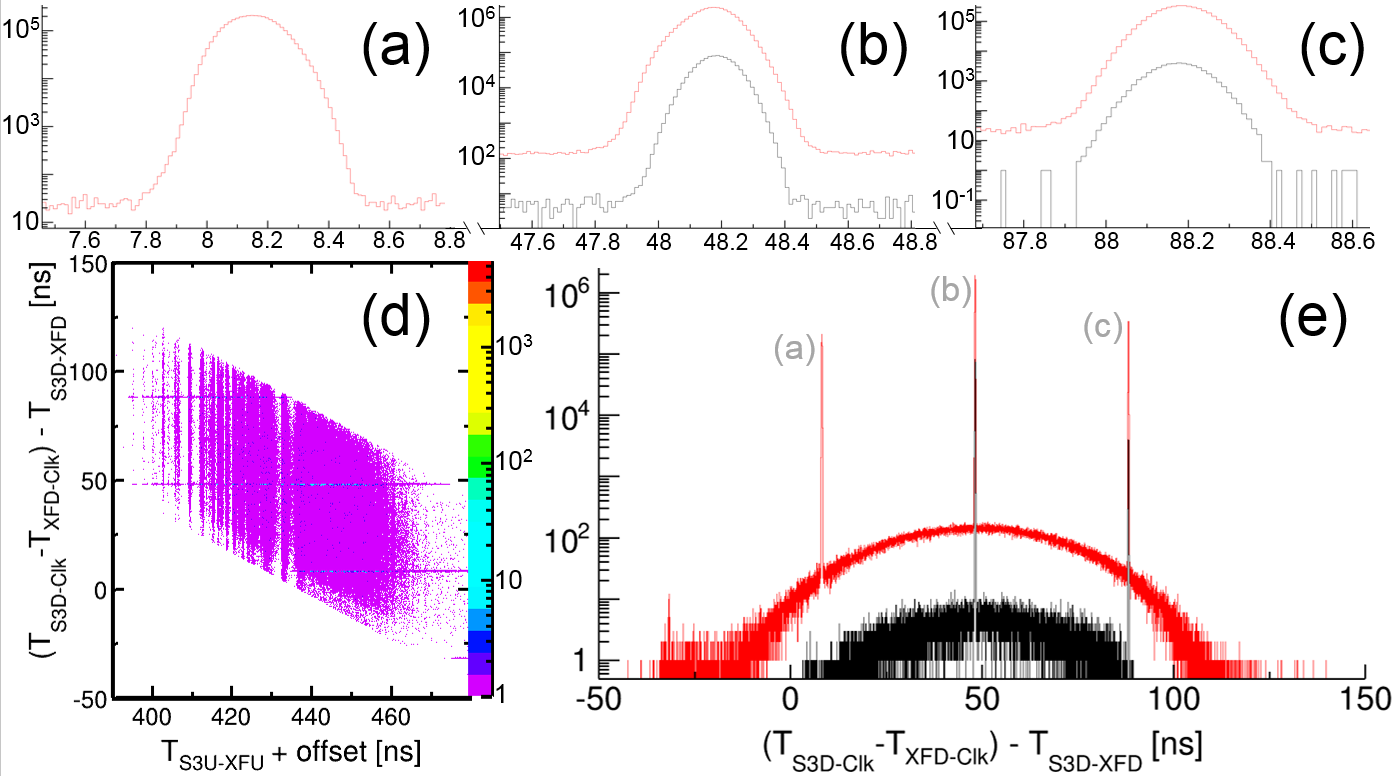}
\caption
{
  (color online.) Spectra employed for the clock pulse correction.
  The time difference between a direct time-of-flight (TOF) and
  clock TOF (panel e), 
  results in multiple peaks
  (panels a--c) spaced by the clock period $T$=40~ns. Narrow gates around the peaks were
  used to remove background and to determine the clock pulse
  correction that was to be added to a given event. Panel d
  demonstrates the fact that events of an ion with a single direct
  TOF could result in multiple clock TOFs. The random coincidences
  which are prominent in panel d are shown in panel e to be a small fraction of
  total events. The vertical structures in panel d are due to the
  fact that ions with similar $A/Z$ had similar TOF, where the
  feature at $\approx432$~ns corresponds to $A/Z=2.5$. The black histograms in
  panels a--c and e are gated on events of $^{45}$Ar, the highest
  statistics isotope observed, while the red
  histograms are for all events.
}
\label{fig:ClockPulseCorrectionDown}
\end{center}
\end{figure*}

Ultimately,
TOF was measured for $\sim$150~nuclides, ranging from atomic number
$14\lesssim Z\lesssim30$ and atomic mass to atomic number (here
the ion charge $q=Z$) ratio $2.35\lessapprox A/Z\lessapprox2.72$.
The measured TOFs were
in the range of $\sim500\pm25$~ns. The event-by-event TOFs
were corrected for their $B\rho$ variation due to the finite $B\rho$
acceptance of the ion optical system using a globally-fit (i.e. fit over the
full range of nuclides) correction based on the measured position at
the S800 target location. The resultant single-species TOF distributions for
the $B\rho$-corrected data were fit with a Gaussian
distribution in order to determine a mean TOF
for each nuclide. The relationship between mass over charge
		  $m_{\rm{rest}}/q$
and TOF was fit to the data of reference nuclides in order to
ascertain the
calibrated $m_{\rm{rest}}/q(\rm{TOF})$ relationship that was used to
obtain the measured masses reported in this work.

\subsection{Timing measurement}
\label{ssec:Timing}

The method employed by Ref.~\cite{Mato12} was used to measure the
TOF for nuclides in the mass measurement reported here. Two
1~cm-tall$\times$1.5~cm-wide$\times$0.25~cm-thick 
BC-418 ultra-fast timing scintillators from
Saint-Gobain Crystals~\cite{SaintGobain} were each coupled to two
Hamamatsu~\cite{Hamamatsu} R4998 1~in-diameter photomultiplier
tubes (PMT) housed in a H6533 assembly (See
Fig.~\ref{fig:Setup}b.). One timing detector was installed in the
focal plane of the A1900 fragment separator, serving as the stop
detector (after including a delay time). The second timing detector
was installed in the focal plane of the S800 spectrograph. This
choice for start and stop signals prevented triggering the data
acquisition system for ions which did not traverse the full flight
path. The signal from each PMT was
split. One signal was used for timing information and the
other signal was used to measure the magnitude of the light output
for position and $Z$ information. To maintain signal quality, timing
signals were transported to the data acquisition electronics via
Belden~\cite{Belden} model 7810A delay cables.
This
set-up provided an intrinsic timing resolution of
$\sim30$~ps~\cite{Mato12}.

Various combinations were made of the four PMT timing signals, 
one each from the `Up' (low-$B\rho$ side) and `Down' (high-$B\rho$
side) PMTs of the A1900 and S800 timing detector set-up, to create
a TOF for each event, the `event TOF', each of which is discussed in detail in Ref.~\cite{Meis15phd}.
The event TOF which was ultimately chosen to minimize the systematic
uncertainty in the final results is the `Down-Clock' TOF of
Ref.~\cite{Meis15phd}. For this event TOF, the high-$B\rho$ PMT
signals from the S800 and A1900 fast-timing scintillators were each
used to start separate channels of a time-to-amplitude converter
(TAC), which each had a stop signal generated by a
clock. Each separate TAC time randomly populated the full-range of an
analog-to-digital converter (ADC), cancelling out systematic
effects from local-nonlinearities in the ADC channel-to-time mapping
that are difficult to characterize and correct. The
random time-component of the event-TOF timing signals was removed by taking the
difference between the two clock times, referred to as
$T_{\rm{S3D-Clk}}$ and $T_{\rm{XFU-Clk}}$ for the S800 and A1900
low-$B\rho$-side PMT vs. clock times, respectively. 
The event TOF constructed from the clock-stopped time difference,
$T_{\rm{XFD-Clk}}-T_{\rm{S3D-Clk}}$, for a
given flight-time could vary by an integer multiple of the clock
period ($T=40$~ns), since the clock pulses came at random intervals with respect
to the ion flight-time measurement. The event TOF was corrected for
the number of clock pulses via a comparison to the direct
time-of-flight measured between the two low-$B\rho$-side PMTs, as
shown in Fig.~\ref{fig:ClockPulseCorrectionDown}. An additional
correction was applied to each event-TOF to account for the
systematic shift associated with an ion's scintillator
impact-positions, which were obtained from the direct
time-difference between the opposing PMTs on each of the fast-timing
scintillators, $T_{\rm{XFU-XFD}}$ and $T_{\rm{S3U-S3D}}$.

The event-by-event TOF for each ion was
\begin{eqnarray}
\rm{TOF}_{\rm{event}}&=&T_{\rm{XFD-Clk}}-T_{\rm{S3D-Clk}}\nonumber\\
&+&N_{\rm{d}}T+\frac{1}{2}(T_{\rm{XFU-XFD}}-T_{\rm{S3U-S3D}})\nonumber\\
&+&t_{\rm{offset}},
\label{eqn:eventTOF}
\end{eqnarray}
where $N_{\rm{d}}$ is the number of clock pulses to correct for (via
Fig.~\ref{fig:ClockPulseCorrectionDown}) and
$t_{\rm{offset}}=$480~ns is an arbitrary offset applied to bring
measured TOFs closer to the expected true TOFs, which differ due to
the chosen delay-cable lengths.

\subsection{Rigidity determination}
\label{ssec:Rigidity}

\begin{figure}[]\begin{center}
\includegraphics[width=1.\columnwidth,angle=0]{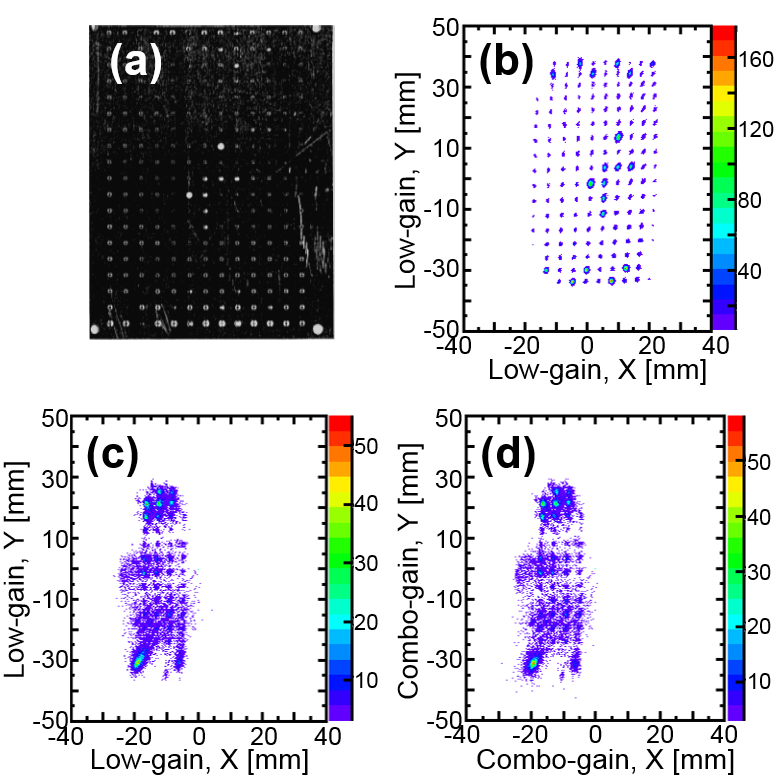}
\caption
{
  (color online.) Panel (a) shows the mask with a distinctive hole pattern (5~mm
  hole-spacing) which was placed in between the incoming ion and
  gold foil in order to only allow electrons to be created from
  certain locations for calibration runs. Panel (b) shows the image
  created on the MCP by electrons generated from a $^{232}$Th
  $\alpha$-source. Panels (c) and (d) show the image created by the
  electrons generated by the $^{82}$Se primary beam, where the beam
  was tuned to four separate positions to achieve the mask-coverage
  shown, where the low-gain corner
  signals were used for panel (c) and the combined high-low gain
  signals were used for panel (d). Since only the relative position
  was relevant, the effort was not made to achieve the exact 5~mm
  hole-spacing of the mask in the MCP image.
}
\label{fig:MCPHoleMask}
\end{center}
\end{figure}

A relative measurement of $B\rho$ was performed using the method
developed by Ref.~\cite{Shap00} at the target
position of the S800 spectrograph, which was operated in a dispersion
matched mode~\cite{Bazi03}. This consisted of sending the ion beam
through a foil and guiding the secondary electrons generated in this
process to the surface of an 8~cm-wide$\times$10~cm-tall (where the
width is along the non-dispersive direction) 
microchannel plate detector (MCP) (See Fig.~\ref{fig:Setup}.). The foil was a 70~$\mu$g$\,$cm$^{-2}$
polypropylene film sputtered with 1500~$\AA$ of gold biased to
$-1$~kV, which provided an electric field to guide electrons
directly from the foil to the MCP, the face of which was at ground
potential. The MCP consisted of two Quantar~\cite{Quantar} model 3398A
lead-glass plates oriented in the chevron configuration.
Rectangular NdFeB 35 permanent magnets from Magnet Sales and
Manufacturing~\cite{MagnetSales} were held co-planar to the foil
and MCP by a steel yoke in order to create a region of nearly homogeneous magnetic field
between the foil and MCP, so that the secondary electrons would
follow a tight spiral along their flight path. The secondary
electrons were multiplied by the MCP in an avalanche which was
collected on a resistive back plane, where electrons freely drifted
to its four corners. The foil was mounted on a ladder which also
contained a foil and hole-mask with a known hole pattern, shown in
Fig.~\ref{fig:MCPHoleMask}a, that was
used for the dispersive position ($\propto B\rho$) calibration.

Ion impact positions on the MCP, and therefore on the foil, were
reconstructed by determining the relative amount of charge collected
on each corner of the resistive back plane. For a single event, the
non-dispersive $X_{\rm{MCP}}$ and dispersive $Y_{\rm{MCP}}$ positions of an ion at the
foil were given by
\begin{eqnarray}
&&
X_{\rm{MCP}}=\frac{\rm{UR}+\rm{LR}-\rm{UL}-\rm{LL}}{\rm{UL}+\rm{UR}+\rm{LL}+\rm{LR}}\nonumber\\
&&
Y_{\rm{MCP}}=\frac{\rm{UL}+\rm{UR}-\rm{LL}-\rm{LR}}{\rm{UL}+\rm{UR}+\rm{LL}+\rm{LR}},
\label{eqn:XYpositions}
\end{eqnarray}
where $\rm{UL}$, $\rm{UR}$, $\rm{LL}$, and $\rm{LR}$ are the charges
collected on the upper left, upper right, lower left, and lower
right corners, respectively, of the MCP back plane. Each corner signal was split
and sent through low and high-gain amplification, which were optimum
for positions close to and far from a given corner, respectively. In practice, the positions reconstructed from the low-gain
amplification were of comparable quality to the
combined-gain positions, as seen in
Fig.~\ref{fig:MCPHoleMask}, and so the low-gain corner
signals were used for the final MCP position determination. The
achieved position resolution was $\sigma\approx0.5$~mm and
$\sigma\approx1.0$~mm for secondary electrons generated by a
$^{228}$Th $\alpha$-source and $^{82}$Se primary beam, respectively,
where the lower resolution for the primary beam was due to the
larger initial kinetic energy of the secondary
electrons~\cite{Jung96},
and therefore larger cyclotron radius~\cite{Land75,Roge15}.

\begin{figure}[]\begin{center}
\includegraphics[width=1.0\columnwidth,angle=0]{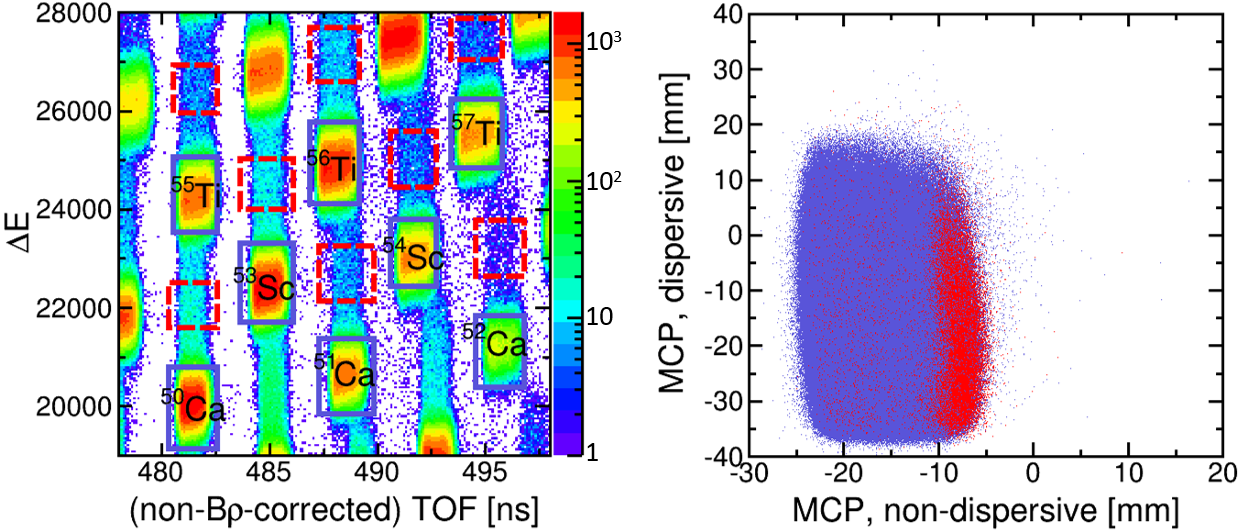}
\caption
{
   (color online.) Demonstration of the correlation between high energy-loss
	($\Delta E$) PID
	events and the microchannel plate (MCP) non-dispersive position. The
	left panel shows a subset of the PID containing isotopes of
	calcium, scandium, and titanium, where `main' events are within
	the purple boxes and `top-hat' events are within the red-dashed
	boxes. The right panel shows the location of the `main' (purple
	dots) and `top-hat' (red dots) events on the MCP, where it is
	clear that the relatively high $\Delta E$
	events corresponded to larger non-dispersive positions.
}
\label{fig:PIDtophat}
\end{center}
\end{figure}

In addition to providing a relative measure of $B\rho$, the MCP
position measurements were used to identify scattering on a
collimator upstream of the foil that was used to protect the MCP
during beam tuning. Scattering on the collimator reduced the energy
of the scattered fragment, resulting in an increased energy loss in
the S800 focal plane ionization chamber that was used for PID. These
scattered events added a `top-hat' feature above the `main'
(non-scattered) events in the PID, as shown in
Fig.~\ref{fig:PIDtophat}. A position gate, $X_{\rm{MCP}}<-11$~mm, was applied
to remove scattered events from the analysis.

\section{Data analysis}
\label{sec:DataAnalysis}

\subsection{Rigidity correction}
\label{ssec:RigidityCorr}

Due to the accepted momentum spread of $\pm0.5$\%,
a rigidity correction was required to remove the
momentum-dependence from the measured TOF spectra. The $B\rho$ correction was first
determined individually for each nuclide, the `local'
$B\rho$-correction, by fitting the TOF-$Y_{\rm{MCP}}$ relationship for the set
of events belonging to a given nuclide. The parameters of the local
rigidity corrections were then fit to determine a smooth variation
of these parameters as a function of $A$ and $Z$, resulting in the
`global' $B\rho$ correction which was ultimately used to momentum-correct the
data. The global correction function allows for the momentum
correction of nuclides with low statistics, for which a precisely
determined local correction was not possible, removes
spurious systematic effects from unphysical variations in the local
rigidity corrections due to limited statistics, and its use for all
nuclides ensures a consistent treatment of the data.

\begin{figure*}[]\begin{center}
\includegraphics[width=2.0\columnwidth,angle=0]{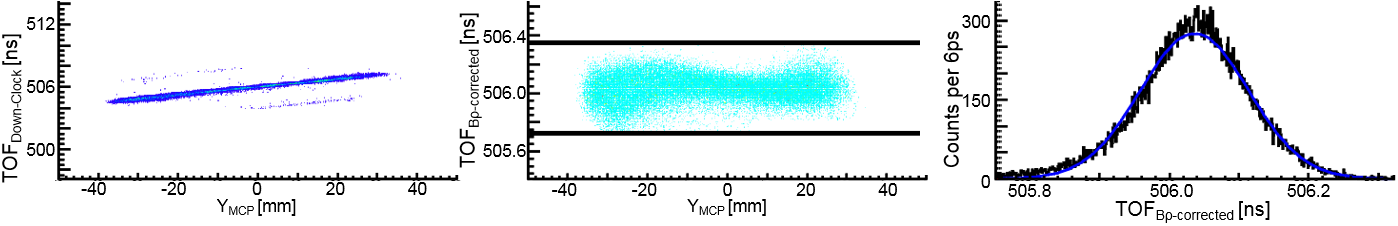}
\caption
{
 (color online.) The first iteration of the local
 $B\rho$-correction for $^{64}$Cr.
 The left
 panel shows a histogram of TOF vs $Y_{\rm{MCP}}$ for events
 identified as $^{64}$Cr, which was converted to a
 graph by applying {\tt ROOT}'s {\tt TProfile} class to the
 histogram and fit with a linear function.
 The
 middle panel shows the resultant $B\rho$-corrected TOF vs
 $Y_{\rm{MCP}}$ histogram after removing the linear trend found in
 the left panel, pivoting about $Y_{\rm{MCP}}$=0.
 The right panel shows the projection onto the TOF-dimension of the
 rigidity corrected (black
 histogram) TOF vs $Y_{\rm{MCP}}$ relationship, where the blue line
 is a Gaussian fit. The black lines in the middle panel indicate
 $\pm4\sigma$, where $\sigma$ is the standard deviation of the
 Gaussian fit of the right panel.
}
\label{fig:LocalBrhoCorrCr64}
\end{center}
\end{figure*}

The local $B\rho$-correction was performed isotope-by-isotope in an
iterative fashion. First, the TOF vs $Y_{\rm{MCP}}$ data for an
isotope were histogrammed,
converted into a graph with ${\tt ROOT}$'s ${\tt
TProfile}$~\cite{ROOT}
class, and fit with a linear function (See
Fig.~\ref{fig:LocalBrhoCorrCr64}). A linear function was chosen as
it was found to
reduce the overall systematic uncertainty in the final
mass-fit~\cite{Meis15phd}.
The linear dependence of TOF on
$Y_{\rm{MCP}}$ was then removed (See
Fig.~\ref{fig:LocalBrhoCorrCr64}b), the data were projected onto the
TOF dimension, and the projected histogram was fit with a normal
distribution (See Fig.~\ref{fig:LocalBrhoCorrCr64}c). Due
to contamination from misidentified nuclei in the PID, the TOF vs
$Y_{\rm{MCP}}$ spectra contained two weak lines parallel to the
main linear data trend, offset to higher and lower TOF, since low $B\rho$ (low TOF) events from
higher-TOF nuclides could be misidentified as high $B\rho$ (high
TOF) events from the nuclide of interest and vice versa for events
from lower-TOF nuclides. The $B\rho$ measurement allowed these
misidentified nuclei to readily be identified in the TOF vs
$Y_{\rm{MCP}}$ spectra, however they skewed the slope of the initial
linear fit.  Therefore, following the fit-correction-projection-fit
procedure shown in Fig.~\ref{fig:LocalBrhoCorrCr64}, a cut was
made to only select events within $4\sigma$ from the TOF centroid of
the normal distribution fit. The fit-correction-projection-fit
procedure was then repeated until convergence was reached to obtain the slope of the local
$B\rho$-correction for that isotope. The linear local $B\rho$-correction
was found to be insufficient for isotopes of elements with $Z<17$ and $Z>26$
and nuclides with $A/Z<2.44$, so these nuclides were excluded
from the analysis. On average the slope of the
TOF--$Y_{\rm{MCP}}$ relationship was $\sim40$~ns$\,$mm$^{-1}$.

\begin{figure*}[]\begin{center}
\includegraphics[width=2.0\columnwidth,angle=0]{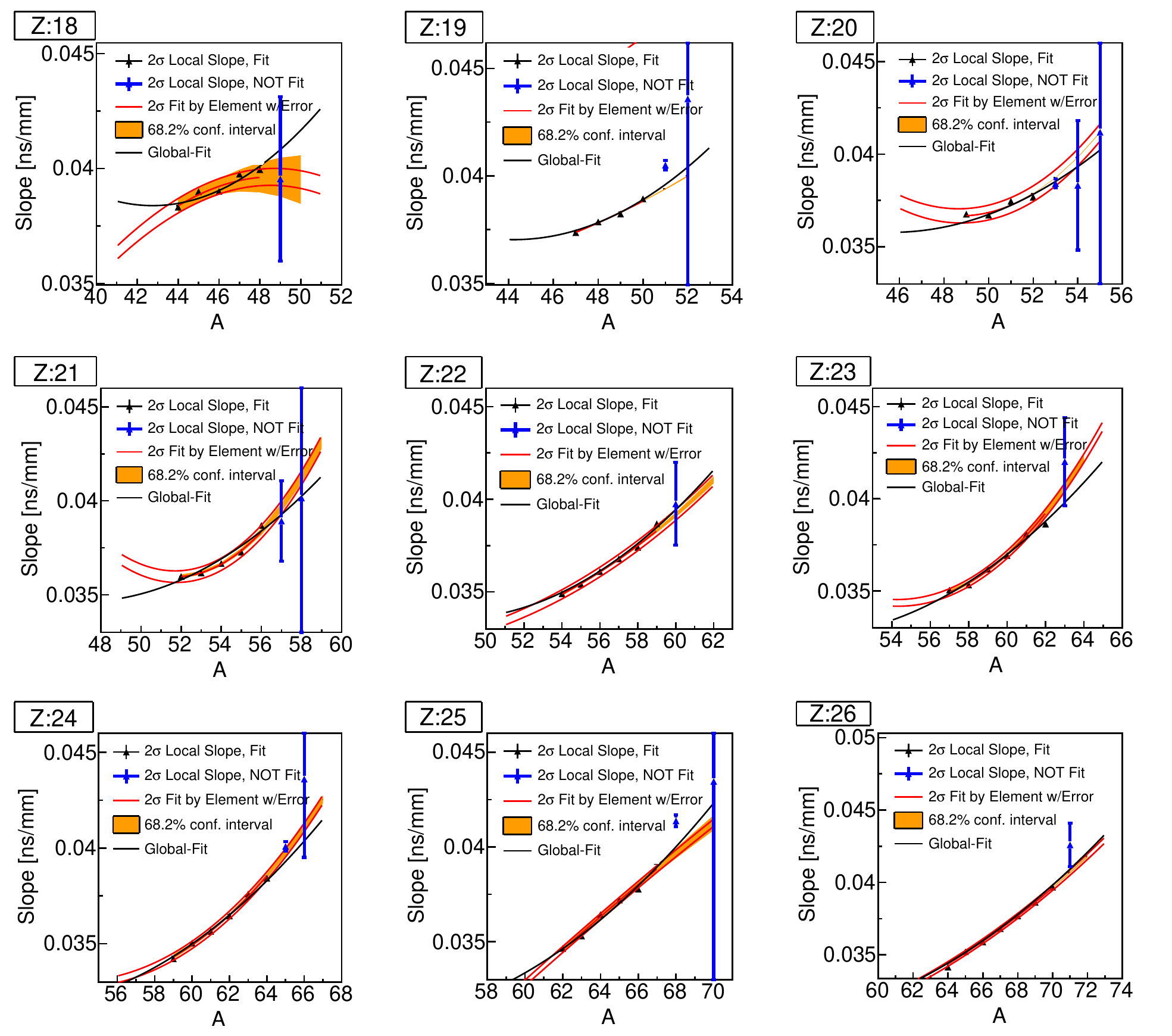}
\caption
{
  (color online.) TOF vs $Y_{\rm{MCP}}$ slope as a function of mass number $A$ for
  observed isotopes of argon, potassium, calcium, scandium,
  titanium, vanadium, chromium, manganese, and iron ($18\leq
  Z\leq26$), respectively, as determined by `local' by-nucleus fits
  (data points) and fits to
  the locally-determined slopes that employed the $\pm2\sigma$
  cut-off, where the black
  data points were included in the fit and the blue points were not.
  The by-element fit along a single isotopic chain
  as a cubic function of $A$ is shown by the red lines, where the
  upper and lower lines indicate the extremes obtained for upper and
  lower limits of the fit-parameters, and the orange band indicates
  the $\pm1\sigma$ confidence interval. The black line shows the
  trend of the rigidity-correction slope along an isotopic chain as determined by the global
  fit to all locally-determined slopes of nuclei with $A/Z>2.44$ and $18\leq Z\leq26$.
}
\label{fig:LocalGlobalSlopeTrends}
\end{center}
\end{figure*}

The locally determined linear dependencies of TOF on
$Y_{\rm{MCP}}$ were then fit to determine a global $B\rho$-correction.
Various polynomials in $A$, $Z$, and $A/Z$ were explored, up to fourth order
in each variable, and the optimum fit-function in terms of goodness
of fit was selected:
\begin{eqnarray}
\left(\frac{d\rm{TOF}}{dY_{\rm{MCP}}}\right)_{\rm{global}}=a_{0}&+&a_{1}\frac{A}{Z}+a_{2}\left(\frac{A}{Z}\right)^2+a_{3}Z\nonumber\\
&+&a_{4}Z^{2}+a_{5}A,
 \label{eqn:GlobalSlopeFitFn}
\end{eqnarray}
where $a_{i}$ are fit parameters. The global $B\rho$-correction
slopes from this fit reproduced the local $B\rho$ correction slopes
within 1\%. The same optimum global fit
function was found by Ref.~\cite{Estr11}. An element-by-element fit
to the local $B\rho$-correction slopes was also explored, though it was
found to be inferior in terms of the final mass-fit systematic
uncertainty~\cite{Meis15phd}. The local, global, and by-element
$B\rho$-correction slopes are compared in
Fig.~\ref{fig:LocalGlobalSlopeTrends}. Note that isotopes with
$Z=17$ are not shown since they were ultimately excluded from the
analysis due to their drastically different behavior in TOF as a
function of $m/q$, as determined by the mass-fit (Recall isotopes of
elements with $Z<17$ and
$Z>26$ were previously excluded from the analysis due to their poor
local $B\rho$ correction determination.).

 \begin{figure}[ht]
 \includegraphics[width=1.0\columnwidth,angle=0]{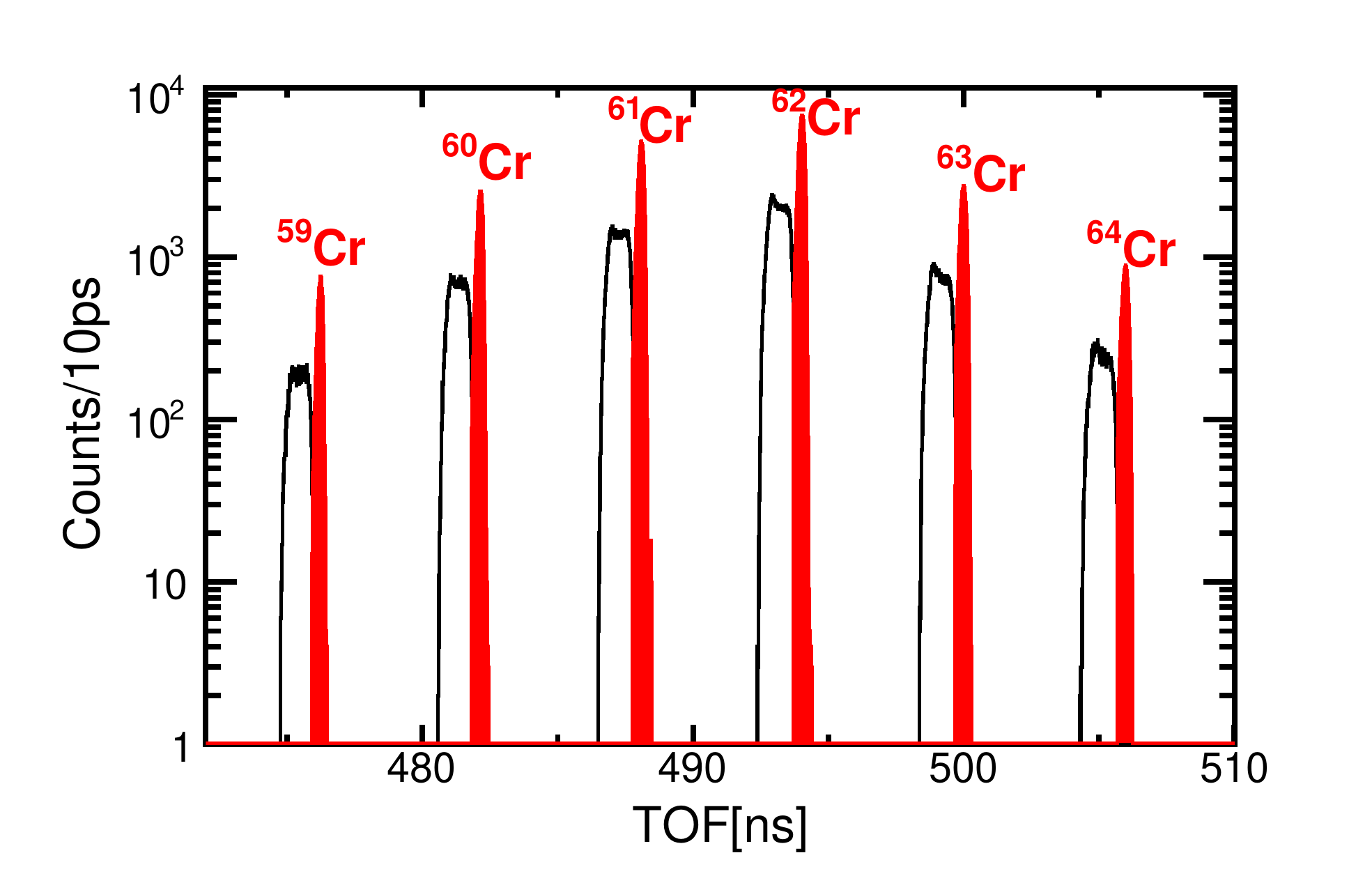}
 \caption{(color online). Time-of-flight (TOF) distribution of
 chromium isotopes before (open histograms) and after (red-filled
 histograms) the global magnetic rigidity correction.
 \label{fig:CrTOFrigiditycorrection}}
 \end{figure}

The global $B\rho$-correction was applied to the TOF spectra, as is
shown in Fig.~\ref{fig:CrTOFrigiditycorrection} for the chromium
isotopes, where it is apparent that a shift in the average TOF of
the distribution occurs due to the choice of $Y_{\rm{MCP}}$ which
TOF was pivoted about. The $B\rho$-correction improved
$\sigma_{\rm{TOF}}$ from $\sim$2~ns to $\sim$0.08~ns. The final TOF
for each nuclide was determined by fitting the $B\rho$-corrected TOF
with a normal distribution, gating on events within $\pm4\sigma$ of
the TOF centroid,
and repeating the fitting-gating procedure until convergence. The
statistical uncertainty of the mean TOF for measured nuclides was
$\delta \rm{TOF}\lesssim1$~ps, corresponding to a TOF measurement
precision of roughly one part in $10^{6}$.

\subsection{Mass evaluation}
\label{ssec:MassEval}

\begin{figure}[]\begin{center}
\includegraphics[width=1.0\columnwidth,angle=0]{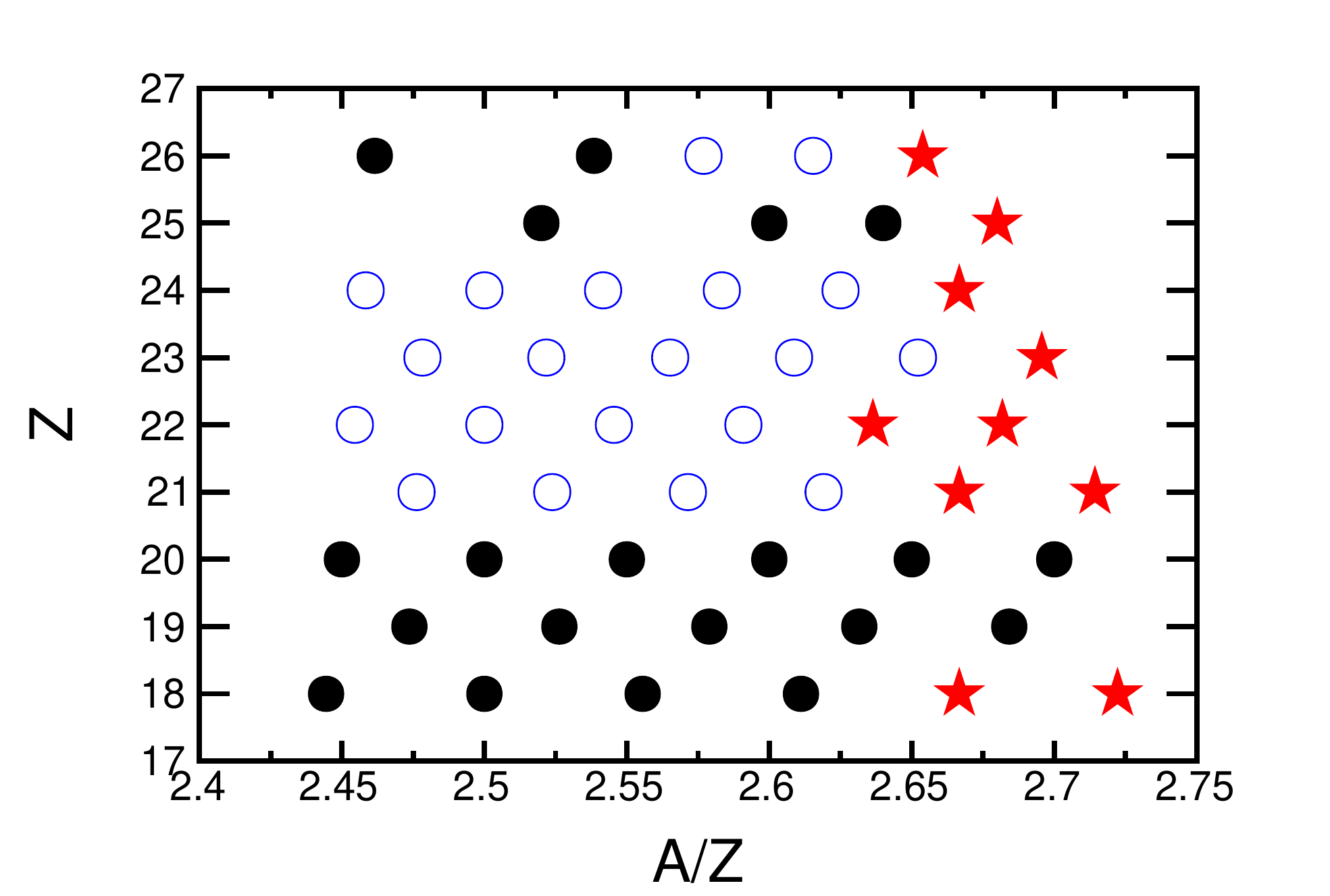}
\caption
{
  (color online.) Map of nuclides observed in the TOF mass
  measurement (with
  sufficient statistics to obtain a TOF value) in terms of atomic
  mass number to nuclear charge ratio $A/Z$ and nuclear charge $Z$.
  Solid black circles indicate reference nuclides, open blue circles
  indicate nuclides with masses known in the literature, but not to
  sufficient precision to qualify as reference nuclides, and red stars
  indicate nuclides with unknown mass prior to this experiment. The
  isotopes $^{63}$Mn and $^{63,65}$Fe were not considered, as they
  have known low-lying isomers that preclude these nuclides as
  calibrants of the mass fit.
  Our results for $Z=18$ and $Z=21$ are published in
  Refs.~\cite{Meis15} and \cite{Meis15b}, respectively. Our results
  for $Z=25,26$ will be the subject of a forthcoming publication.
}
\label{fig:ReferenceMassMap}
\end{center}
\end{figure}

The fit to the mass over charge $m/q$--TOF surface, elsewhere
referred to in this article as the `mass-fit', consisted of choosing a set of
reference nuclides to calibrate the $m_{\rm{rest}}/q(\rm{TOF})$
relationship, finding the optimum fit function, and assessing the
various uncertainties contributing to the final mass results
obtained for nuclides that were not used as calibrants. Nuclides
chosen as calibrants had a literature experimental mass uncertainty
$\leq50$~keV, as listed in the 2012 Atomic Mass
Evaluation~\cite{Audi12} (except for $^{53}$Ca and $^{54}$Ca which
come from Ref.~\cite{Wien13}), and no isomers longer-lived than
100~ns, as listed in the National Nuclear Data Center
database~\cite{NNDC}. The twenty nuclides used to calibrate the
$m_{\rm{rest}}/q(\rm{TOF})$ relationship were $^{44-47}$Ar,
$^{47-51}$K, $^{49-54}$Ca, $^{63,65,66}$Mn, and $^{64,66}$Fe. A map
of the reference nuclides with respect to the nuclides for which a
mass was evaluated is shown in Fig.~\ref{fig:ReferenceMassMap}.

The atomic masses from Ref.~\cite{Audi12} were corrected to obtain
nuclear masses by subtracting the individual electron binding energies listed
in Table II of Ref.~\cite{Lotz70}. A relativistic correction was applied to the measured TOF for
nuclides in order to account for time-dilation. Additionally, the
average TOF and $Z$ for all nuclides of interest were subtracted from
the TOF and $Z$ of each nuclide to create effective time and charge
variables, i.e. $\tau=\rm{TOF}-\langle\rm{TOF}\rangle$ and
$z=Z-\langle Z\rangle$, in order to reduce the multicollinearity of
the mass-fit parameters~\cite{Mato12}.

The initial uncertainty in $m_{\rm{rest}}/q$ ascribed to the data points
was the literature mass uncertainty added in quadrature to
the statistical uncertainty, where the latter used standard
propagation of uncertainty to translate uncertainty in TOF into
uncertainty in $m/q$.
This statistical uncertainty
depended on the fit function itself, $\delta
M_{\rm{stat.}}=\left(\delta\rm{TOF}\right)\times\frac{\partial}{\partial\rm{TOF}}\left(\frac{m}{q}(\rm{TOF})\right)$
where $\frac{m}{q}(\rm{TOF})$ is the $m_{\rm{rest}}/q(\rm{TOF})$
calibration
function
and $\delta\rm{TOF}$ is the one standard deviation uncertainty of
the mean TOF for a nuclide (data point). Therefore, the final statistical
uncertainty assigned to each data point was determined in an iterative
procedure where the data was fit to obtain a
$m_{\rm{rest}}/q(\rm{TOF})$
calibration
function, statistical uncertainties were calculated for each of the
data-points (corresponding to reference nuclides), and the process was
repeated until convergence.

\begin{figure}[]\begin{center}
\includegraphics[width=1.0\columnwidth,angle=0]{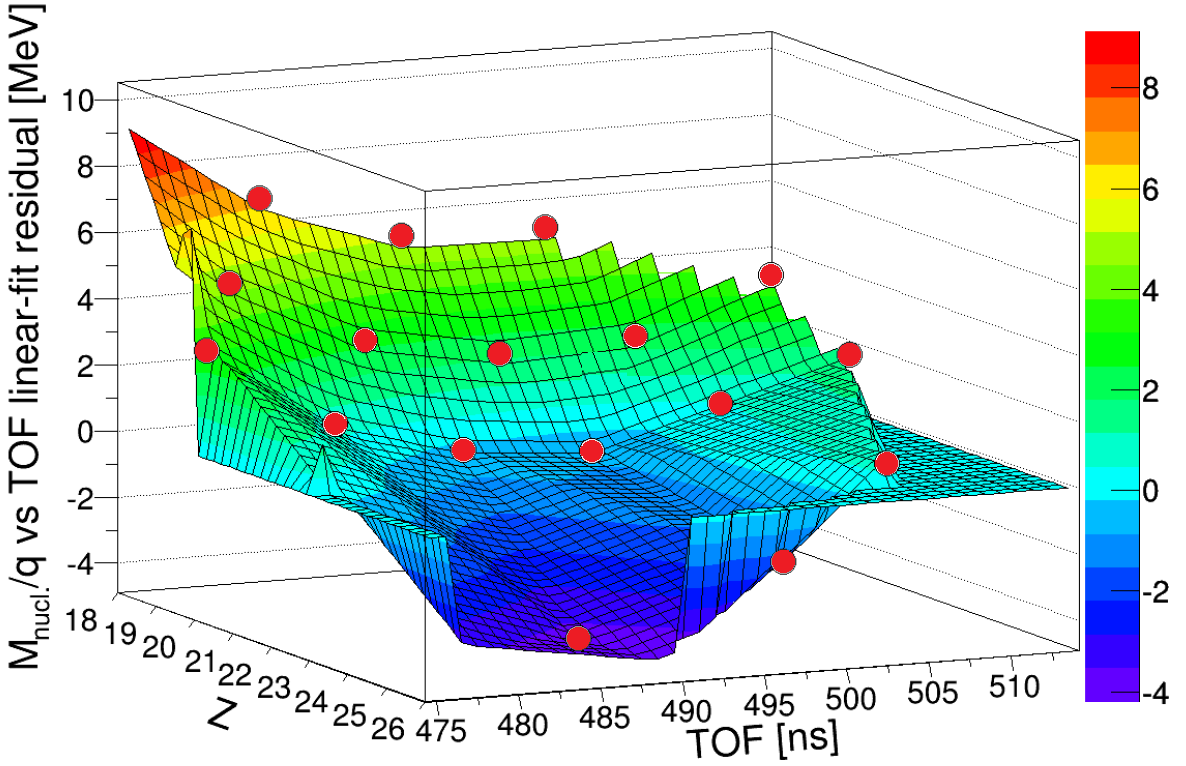}
\caption
{
  (color online.) $m_{\rm{rest}}/q$--TOF surface of reference
  nuclides where the linear dependence
  of mass over charge $m/q$ on TOF has been removed. Solid red
  points mark the nuclear charge $Z$ and TOF of reference nuclides
  while the color of the surface at that location indicates the
  linear fit residual in MeV. (Note that the flat region that is
  present
  outside of the region bounded by data points is a feature of the
  plotting software.)
}
\label{fig:MassTOFsurface}
\end{center}
\end{figure}

Upon completion of the mass-fit, including literature and statistical
uncertainties, the reduced $\chi^{2}$ of the fit was typically much
larger than one. This indicated that the uncertainty of the twenty
reference nuclide data-points was underestimated and that some
additional heretofore unaccounted for uncertainty was present. As
there were no systematic trends in the fit residuals, we treat the
additional uncertainty as a systematic error. The
approach outlined in \cite{Mato12} was followed, where the missing
uncertainty was treated as a statistically-distributed systematic
uncertainty, i.e. one that accounted for a uniform scatter in the mass-fit
residual as a function of $m_{\rm{rest}}/q$ (We note that a similar
procedure has been used previously in storage ring isochronous mass
spectrometry~\cite{Chen12}.).
Such an effect could have been
created by many uncontrolled factors in the measurement, such as
time-dependent magnetic field drift of the dipole magnets along the
beam line, time-dependent variations in the response of the timing
electronics due to variations in temperature, or unidentified biases
present in the data analysis pipeline. To
include this additional systematic uncertainty,
the uncertainty of reference nuclide data-points
was increased uniformly, i.e. each data point had the same systematic uncertainty $\delta
M_{\rm{syst.}}$ (in keV$\,q^{-1}$), until $\chi^{2}_{\rm{red.}}=1$.
We note that the
results of the mass-fit with and without inclusion of the systematic
uncertainty agreed within the final one standard deviation
uncertainty.
The mass-fit was then repeated
and the statistical uncertainty was recalculated to be consistent
with the current parameters of the fit function. 
This process was then
repeated iteratively until it converged. The fit-function
resulting from this procedure was the $m_{\rm{rest}}/q(\rm{TOF})$ calibration function
which was used to obtain masses for non-calibration nuclides
whose TOF was measured.

Since the relationship between mass and TOF at the precision level
required to make a meaningful mass measurement was a priori unknown,
several fit
functions were tried, each of which was a combination of polynomials
in TOF, nuclear charge $Z$, and/or a combination of these variables.
The goal of this approach was to find the minimum number of terms
that reproduce the calibration mass surface without any systematic
trends in the fit residuals. This ensures maximum robustness against
interpolation and small-distance extrapolation.
The complex nature of the $m_{\rm{rest}}/q$--TOF surface (See
Fig.~\ref{fig:MassTOFsurface}.) clearly necessitated higher orders
in both TOF and $Z$. A step-by-step procedure was taken
to justify the inclusion of each term added to the mass-fit
function. To be included in the fit function, an extra term
had to significantly reduce the fit residuals and not introduce any
systematic trends. The final mass-fit function which was chosen
represents the minimal set of terms that minimizes the overall
residual to literature masses of the twenty reference nuclides and
resulted in no detectable systematic biases (i.e. trends in the
mass-fit residuals).
As might be expected, some degeneracy existed as to the
benefit of including certain terms in the fit-function. This set of
`best' fits was used to inform the uncertainty of masses evaluated
from the mass-fit function present from extrapolation-from and
interpolation-between the mass-fit calibration points (See
Section~\ref{ssec:MassUncertainty}.). 

\begin{figure}[]\begin{center}
\includegraphics[width=1.0\columnwidth,angle=0]{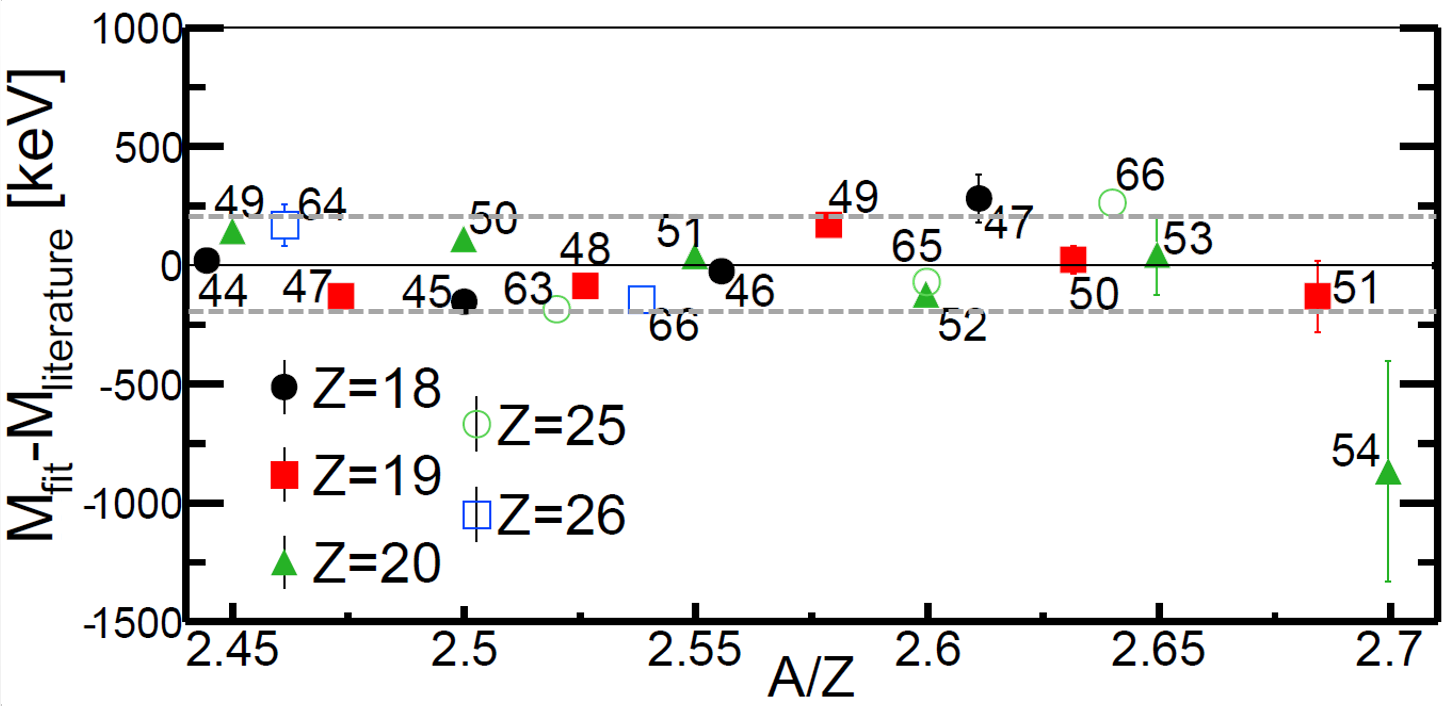}
\caption
{
(color online.) Residuals of the fit to the $m/q(\rm{TOF})$
relationship of calibration nuclides ($^{44-47}$Ar, $^{47-51}$K,
$^{49-54}$Ca, $^{63,65,66}$Mn, and $^{64,66}$Fe) as a function of
the mass number to nuclear charge ratio $A/Z$. Error bars indicate
statistical uncertainties only. The horizontal dashed-gray lines
indicate the average systematic mass uncertainty (9~keV/$q$)
included for reference nuclides for the mass fit, as described in
Section~\ref{ssec:MassEval}.
}
\label{fig:OurResid}
\end{center}
\end{figure}

The final
mass-fit function employed for the mass results was
\begin{equation}
\frac{m}{q}\left(\tau\right)=a_{0}+a_{1}\tau+a_{2}z+a_{3}\tau^{2}+a_{4}z^{2}+a_{5}z\tau+a_{6}z^{4},
\label{eqn:FinalFitFunction}
\end{equation}
where $a_{i}$ are fit parameters.
The optimum mass-fit function (of the set explored) and
the mass results obtained with Eqn.~\ref{eqn:FinalFitFunction}
were found to be robust with respect to the removal of a subset of
reference nuclides from the mass-fit~\cite{Meis15phd}.
Fig.~\ref{fig:MassTOFsurface} shows
Eqn.~\ref{eqn:FinalFitFunction} fit to the
$m_{\rm{rest}}/q(\rm{TOF})$ data for calibration nuclides. The mass-fit
residuals for Eqn.~\ref{eqn:FinalFitFunction} are shown in
Fig.~\ref{fig:OurResid}.

Eqn.~\ref{eqn:FinalFitFunction} contains one extra term,
$z\tau$, and favors $z^{4}$ over $z^{3}$ behavior with respect to
the previous TOF mass measurement at the NSCL~\cite{Estr11}. The
$z^{4}$ term is only slightly favored over the $z^{3}$ term, and a
function using the $z^{3}$ term instead is included in the set of
best-fit functions used to evaluate the extrapolation uncertainty
(See Section~\ref{ssec:MassUncertainty}.).
We surmise that the inclusion of the $z\tau$ term is required due
to the extra energy loss induced by the wedge degrader at the
intermediate image of the A1900, which was not present in
Ref.~\cite{Estr11}. This is because $z\tau\propto A$ and, for fixed
$B\rho$, energy loss $\Delta E\propto A$ (since $\Delta E\propto
Z^{2}/E$, $E\propto Av^{2}$, and
$(B\rho)^{2}=(p/q)^{2}\propto((Av)/Z)^{2}=$constant, where $v$ is the ion velocity and $q=Z$ for the fully-stripped ions measured here).

\subsection{Measurement uncertainty}
\label{ssec:MassUncertainty}

The mass uncertainty for measured nuclides which were not reference
nuclides was comprised of a statistical uncertainty determined from
the nuclide's individual count rate, the statistically-distributed
systematic uncertainty which was determined to be present for
reference nuclides (and therefore assumed to be present for
evaluated nuclides), and 
two additional uncertainties were included to account for
the uncertainty in the mass-fit function.
Namely, these 
were the uncertainties of the fit coefficients that
were a result of the uncertainties in the calibration mass values
and TOFs, referred to here as the `Monte Carlo' uncertainty (motivated by the
way it was calculated), and the uncertainty from the choice of the
general form of the fit function, referred to here as the `function
choice'
uncertainty. 

For the Monte Carlo uncertainty assessment, the mass
of each reference nuclide was perturbed by a random amount
commensurate with its uncertainty, the mass-fit was
performed, the fit results were recorded in a histogram, and this
perturbation-fit-histogram procedure was repeated 10,000 times. The
Monte Carlo uncertainties are the
standard deviations of the fit-result mass distributions. 

The function choice uncertainty was defined as the difference between
the highest and lowest mass value for a given nuclide resulting from
the set of mass-fits that were explored which required a systematic
uncertainty less than three times that of the best mass-fit to
produce a reduced $\chi^{2}$ equal to one and showed no systematic
trend in mass-fit residuals. The five fits with six, seven, or eight
parameters which were considered for
the function choice uncertainty were
Eqn.~\ref{eqn:FinalFitFunction} and similar functions which
contained a $z^{3}$ term rather than a $z^{4}$ term, lacked
the $a_{6}$ term altogether, included an additional term that
depended on $\tau^{4}$, and included an additional term that instead
depended on $z$*$\tau^{2}$. The required statistically-distributed
systematic uncertainty required for each of these fit functions was
9.0~keV/$q$, 11.2~keV/$q$, 22.7~keV/$q$, 8.5~keV/$q$, and
8.2~keV/$q$. Note that the eight-parameter mass-fit functions were
not used in lieu of Eqn.~\ref{eqn:FinalFitFunction} as they did not
yield a significant reduction in the required systematic uncertainty
and thus did not contain the minimal set of terms required to
minimize the overall residual to literature masses of the reference
nuclides.

\begin{figure*}[]\begin{center}
\includegraphics[width=2.0\columnwidth,angle=0]{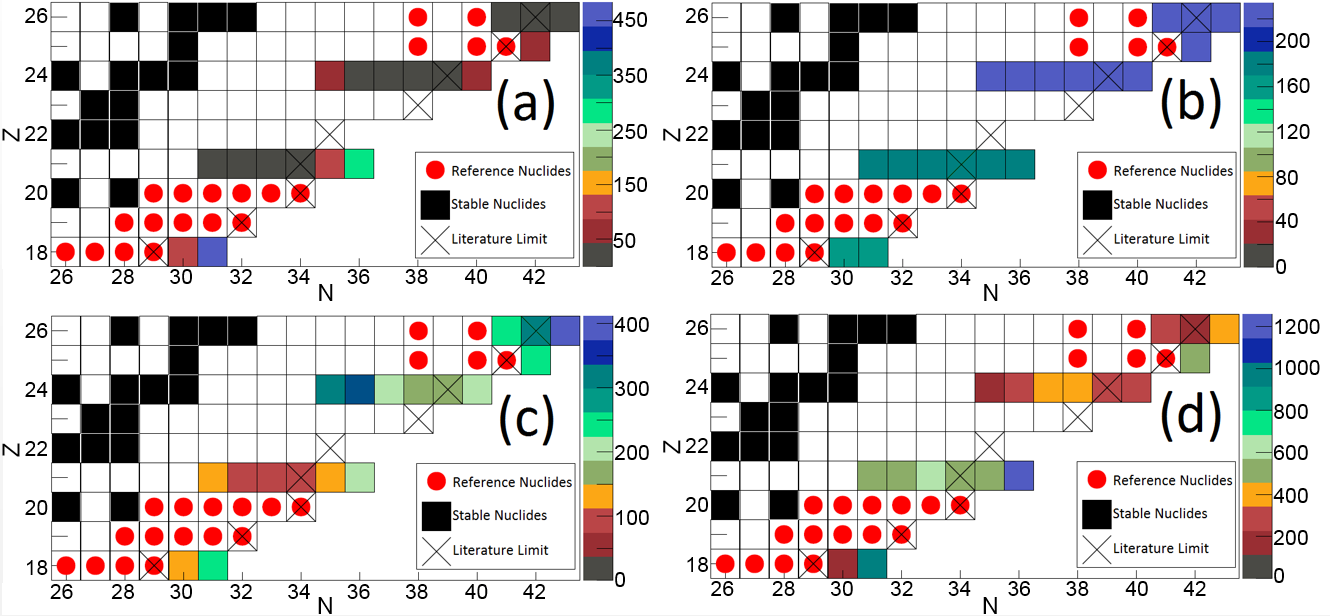}
\caption
{
   (color online.) Statistical (a), systematic (b), Monte Carlo (c),
	and function choice
	(d) uncertainties in keV for nuclides whose mass was
	evaluated in this time-of-flight mass measurement. Colored boxes
	indicate nuclides whose mass was evaluated, with the color
	reflecting the uncertainty in keV, boxes with red circles
	indicate reference nuclides used as calibrants for the
	$m_{\rm{rest}}/q(\rm{TOF})$
	relationship, boxes with
	$\times$'s indicate the most exotic isotope for that element with
	a known mass prior to this experiment, and the black boxes
	indicate stable nuclides.
}
\label{fig:AllUncertainties}
\end{center}
\end{figure*}

\begin{figure}[]\begin{center}
\includegraphics[width=1.0\columnwidth,angle=0]{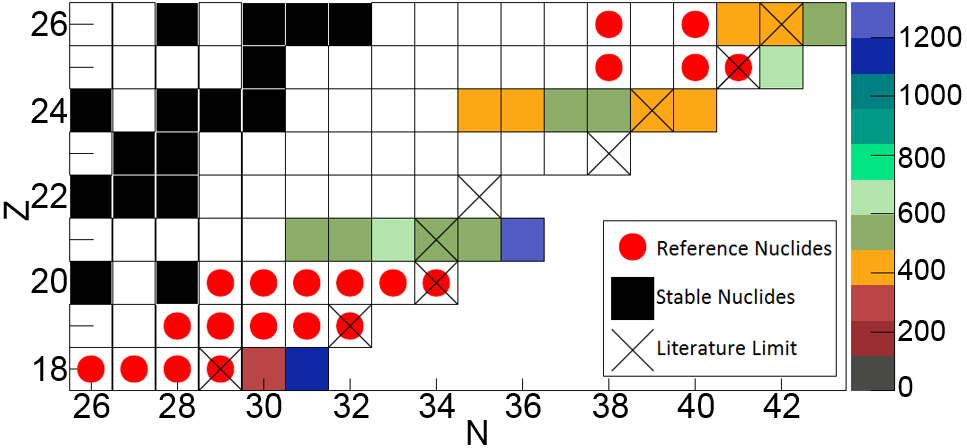}
\caption
{
    (color online.) Same as panels in Fig.~\ref{fig:AllUncertainties}, but with the
	 color indicating the total uncertainty of evaluated nuclide in
	 keV, where the total is the sum in quadrature of the
	 statistical, systematic, Monte Carlo, and function choice
	 uncertainties. Note that $^{56}$Sc has an additional systematic
	 uncertainty due to the presence of a $\beta$-decaying isomer
	 (See Ref.~\cite{Meis15b} for more detail.) which is not included in this figure.
}
\label{fig:FinalUncertaintyTot}
\end{center}
\end{figure}

Fig.~\ref{fig:AllUncertainties} shows the statistical (a),
systematic (b), Monte Carlo (c), and function choice (d)
uncertainties of the masses evaluated in this experiment. Their sum
in quadrature is shown in Fig.~\ref{fig:FinalUncertaintyTot}. It
is apparent that the relative contribution of the uncertainties
resulting from the mass-fit extrapolation and interpolation, i.e.
the Monte Carlo and function choice uncertainties, becomes larger as
the distance in $m/q$ and $Z$ from reference nuclides increases. For
the chromium isotopes, which are the focus of this work, the function
choice uncertainty dominates, as the $Z$-dependence of the
$m_{\rm{rest}}/q(\rm{TOF})$ relationship is poorly constrained by
the available reference nuclides. New high-precision mass
measurements of neutron-rich isotopes of scandium and vanadium
would improve this situation.

\section{Results}
\label{sec:Results}

\begin{table*}[t]
  \caption{\label{CrMassComparison}
  Atomic mass excesses (in keV) of chromium isotopes measured in
  this experiment compared to results from previous direct mass
  measurements from the Time-of-flight Isochronous (TOFI) spectrometer (TOFI1~\cite{Tu90},
  TOFI2~\cite{Seif94}, and TOFI3~\cite{Bai98}), the adopted value in the
  2012 Atomic Mass
  Evaluation (AME)~\cite{Audi12} (`E' are extrapolations), and predictions from global mass models
  (FRDM~\cite{Moll95} and HFB-21~\cite{Gori10}).}
  \def\arraystretch{1.25}
  \begin{ruledtabular}
  \begin{tabular}{lccccccc}
  Isotope & This experiment & AME 2012 & TOFI1 & TOFI2 & TOFI3 & FRDM &
  HFB-21 \\ \hline
  $^{59}$Cr & $-48~540~(440)$ & $-47~891~(244)$ & $-47~710~(230)$ &
  $-47~850~(250)$ & $-47~320~(310)$ & $-48~680$ & $-49~160$ \\
  $^{60}$Cr & $-47~440~(460)$ & $-46~504~(213)$ & $-46~280~(230)$ &
  $-46~830~(260)$ & $-46~510~(280)$ & $-47~910$ & $-48~200$ \\ 
  $^{61}$Cr & $-43~080~(510)$ & $-42~455~(129)$ & $-41~500~(400)$ &
  $-42~770~(280)$ & $-42~120~(280)$ & $-42~700$ & $-43~710$ \\
  $^{62}$Cr & $-40~890~(490)$ & $-40~895~(148)$ & $-39~500~(600)$ &
  $-41~200~(400)$ & $-40~200~(350)$ & $-41~180$ & $-41~960$ \\
  $^{63}$Cr & $-35~940~(430)$ & $-35~722~(459)$ & \dots & \dots &
  \dots & $-36~030$ & $-37~290$ \\
  $^{64}$Cr & $-33~480~(440)$ & $-33~459\rm{E}~(298\rm{E})$ & \dots
  & \dots & \dots & $-34~950$ & $-34~730$ \\
  \end{tabular}
  \end{ruledtabular}
\end{table*}

The atomic mass excesses for the chromium isotopes measured in this
experiment are compared to theoretical and literature values in
Tab.~\ref{CrMassComparison}, where we note that the mass of
$^{64}$Cr was measured for the first time. These results correspond
to a mass measurement precision of roughly one part in $10^{5}$.

For our mass comparison in Tab.~\ref{CrMassComparison} we 
focus on previous experimental values
reported~\cite{Tu90,Seif94,Bai98} by the Time-of-flight
Isochronous Spectrometer (TOFI) facility, as these results constitute the
primary contribution to the evaluated mass reported for these
isotopes in the 2012 Atomic Mass Evaluation (AME)~\cite{Audi12}. We
compare to the theoretical results reported by the 1995 Finite Range
Droplet Model (FRDM)~\cite{Moll95} and Hartree-Fock-Bogoliubov-21
(HFB-21)~\cite{Gori10} since these models are commonly used in
astrophysics calculations when experimental data are not available
(e.g. Refs.~\cite{Gupt07,Estr11,Pear11,Scha14}). Additionally,
we
compare our mass-differences to those calculated via the shell-model
using different interactions and model spaces. 

Fig.~\ref{fig:CrS2nE2plusTrends} compares the trend in two-neutron
separation energy $S_{2n}$,
$S_{2n}(Z,A)=2\times\rm{ME}_{\rm{neutron}}+\rm{ME}(Z,A-2)-\rm{ME}(Z,A)$,
for
neutron-rich isotopes of chromium determined from masses reported in
this work to the trends obtained for masses from the 2012
AME~\cite{Audi12} and binding energies calculated by the shell-model
employing the GXPF1A Hamiltonian~\cite{Honm05} in the $fp$-shell
model space, as well as shell-model calculations employing a
modified version of the Hamiltonian from Ref.~\cite{Lenz10}, which
is discussed further in the following section.
We note that we extend the $S_{2n}$ trend for the chromium isotopes to $N=40$ for
the first time. 
The energies of the yrast $2^{+}$ excited states are included in
Fig.~\ref{fig:CrS2nE2plusTrends} for
comparison, as this trend conveys similar information regarding the
evolution of nuclear structure along the chromium isotopic chain~\cite{Meis15}.

The discrepancies
in experimentally-based $S_{2n}$ values, which are largest at $N=36$
and $N=38$, primarily stem from the $\sim650$~keV, $\sim950$~keV,
and $\sim600$~keV differences between this work and the AME values
for $^{59}$Cr, $^{60}$Cr, and $^{61}$Cr, respectively. In
particular, the difference between our $^{60}$Cr mass excess
and the adopted AME value causes the $S_{2n}$ trend for
$N=36-38$ to pivot about $N=37$. As seen in Tab.~I of
Ref.~\cite{Audi12}, the 2012 AME values for these three nuclides are
primarily based on three separate measurements from the TOFI
facility~\cite{Tu90,Seif94,Bai98}, amongst which there is a
$\sim500$~keV discrepancy for the reported masses of $^{59,60}$Cr
and a $\sim1700$~keV discrepancy for the reported $^{61}$Cr masses
(See Tab.~\ref{CrMassComparison}.).

\section{Discussion}
\label{sec:Discussion}

\subsection{Structural evolution of the neutron-rich chromium isotopes}
\label{ssec:CrStruct}

 \begin{figure}[ht]
 \includegraphics[width=1.0\columnwidth,angle=0]{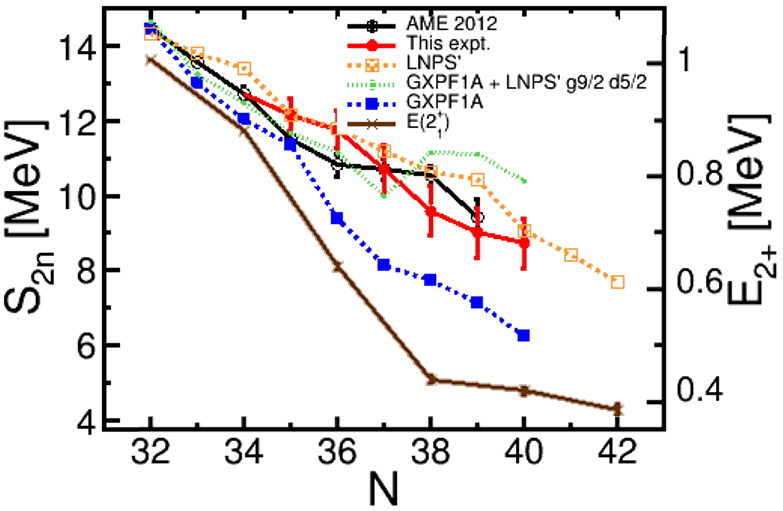}
 \caption{(color online.)
 Two-neutron separation energy $S_{2n}$ for
 neutron-rich isotopes of chromium as calculated from the 2012
 Atomic Mass Evaluation (open black circles) and the masses reported
 here (solid red circles), as well as calculated by shell-model
 calculations employing the GXPF1A Hamiltonian~\cite{Honm05}(solid
 blue triangles) and LNPS' Hamiltonian (modified from
 Ref.~\cite{Lenz10}) (open orange squares). The contribution of the
 $g_{9/2}$ and $d_{5/2}$ orbitals
 is shown by adding their contribution to the LNPS' results to $S_{2n}$
 calculated with the GXPF1A Hamiltonian (green points).  The
 energies of yrast $2^{+}$ excited states of corresponding isotopes
 are shown for comparison (brown
 crosses)~\cite{Rama01,Marg06,Gade10,Wern16}.
 \label{fig:CrS2nE2plusTrends}}
 \end{figure}

The trend in binding energies determined in this work can be used as
a probe of the evolution of shell structure for neutron-rich
chromium isotopes~\cite{Lunn03}. Typically, 
$S_{2n}$ is employed to isolate
the structural changes present along neutron-rich isotopes of an
element~(e.g. Refs.~\cite{Naim12,Gaud12,Wien13,Meis15,Rose15}). Along an isotopic
chain, $S_{2n}$ generally declines with increasing $N$ away from
$\beta$-stability due to the
penalty in binding energy for a large neutron-proton asymmetry, as
described by the liquid-drop model. This decline is markedly
increased following a nucleus that exhibits a magic neutron number,
since the two-neutron removal (addition) required to move from (to)
a nucleus with magic $N$ is energetically disfavored (favored) due
to the shell-gap associated with $N_{\rm{magic}}$~\cite{Lunn03}. A
leveling of $S_{2n}$ for a few isotopes, followed by a continuation
of the gradually decreasing trend is a signature of a shape
transition along an isotopic chain~\cite{Iach11}.

The $S_{2n}$ trends in Fig.~\ref{fig:CrS2nE2plusTrends} demonstrate the different
structural changes implied by the masses presented in this work and
the evaluated masses of the 2012 AME~\cite{Audi12}. Our new data
disfavors the change in the $S_{2n}$ slope at $N=36$ shown by the
2012 AME data, instead favoring a continuation of the previous
slope until $N=38$.
We note that the flattening of the AME $S_{2n}$ trend about $N=36$
is more consistent with the identification of $^{60}$Cr as the
shape-transition point by Ref.~\cite{Brau15}.
The decrease in the magnitude of our $S_{2n}$-trend
slope approaching $N=40$
 is consistent with the collective behavior previously
identified by Refs.~\cite{Gade10,Baug12,Naim12,Craw13,Brau15}. It is
interesting to note that our
$S_{2n}$ trend for $^{62-64}$Cr ($N=38-40$) resembles the trend for
$^{30-32}$Mg~\cite{Chau13} ($N=18-20$), where $^{32}$Mg marks the
entrance of the magnesium isotopic chain into the $N=20$ island of
inversion~\cite{Prit99,Terr08,Wimm10}. However, the masses of chromium isotopes
with $N>40$ are required to provide a firm signature of the presence
or absence of the $N=40$ sub-shell gap for this element.

The striking divergence between the experimental $S_{2n}$ trends and
the shell-model derived trend (GXPF1A) shown in
Fig.~\ref{fig:CrS2nE2plusTrends} highlights the need for inclusion
of the $g_{9/2}$ and $d_{5/2}$ orbits in order to obtain a realistic
description of the chromium isotopes for $N\geq35$, which has been
pointed-out by previous studies~\cite{Deac05,Zhu06,Toga15}. 
We have thus performed large scale shell-model calculations within the 
proton $fp$ and neutron $fpg_{9/2}d_{5/2}$ model space, employing the Hamiltonian 
from Ref. \cite{Lenz10} with minor modifications \cite{Sahin2015, Morfouace15}.
Additionally, the global monopole term was made more attractive by
$30$~keV
to obtain a better agreement of the $S_{2n}$ energies in
neutron-rich chromium and iron isotopes. 
These refinements preserve the spectroscopy of the nuclides in the island of inversion
region presented previously in Ref. \cite{Lenz10}.  

The results of the calculations
with the modified LNPS Hamiltonian, dubbed hereafter LNPS', are also
presented in Fig. \ref{fig:CrS2nE2plusTrends}.
As can be seen, the agreement is more satisfactory than for the
GXPF1A Hamiltonian
and the LNPS' results match with the present data within the error bars for the majority of cases.
The largest discrepancy is found for the $S_{2n}$ value of
$^{63}$Cr,
which is overestimated. This is surprising as the present
model accurately reproduces the known excitation energies of
chromium isotopes, with the visible drop of the yrast $2^+$ excited
state energies
between $N=36$ and $N=38$, indicating that chromium isotopes undergo
a shape change at $N=38$.
However, little is known about the spectroscopy of
$^{63}$Cr~\cite{Such14} and
the ground-state spin assignments of both
$^{63}$Cr and $^{61}$Cr are tentative, making it difficult to
evaluate whether these nuclides
have the correct degree of collectivity in the present shell-model
calculations. 
In spite of this discrepancy, the LNPS' shell-model trend points clearly to the development of collectivity 
around $N=40$ and predicts continuation of the deformation onset
towards higher neutron numbers. This increase in collectivity agrees
with the recent measurement of the yrast  $2^+$ excited state energy
for $^{66}$Cr~\cite{Wern16}.
 
 \begin{figure*}[ht]
 \includegraphics[width=2.0\columnwidth,angle=0]{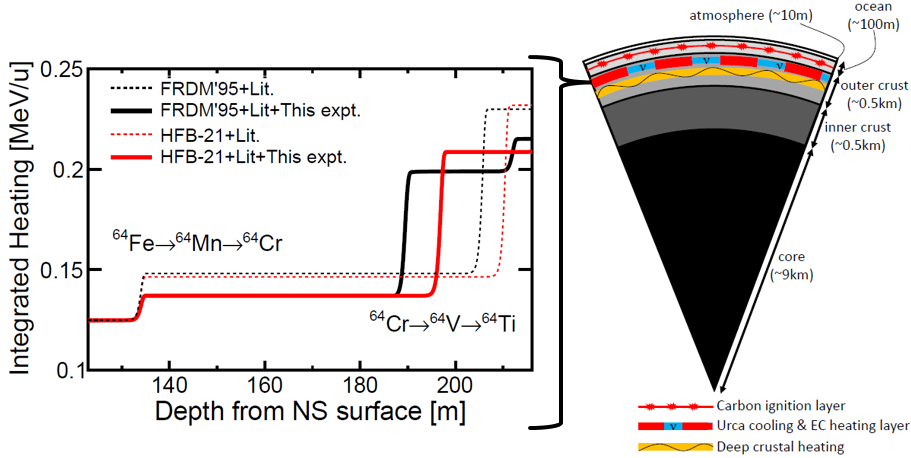}
 \caption{(color online.) Integrated heat release in MeV per
 accreted nucleon from electron capture for an $A=64$ fluid element
 as a function of depth (left panel) in the region where the
 $^{64}\rm{Fe}\rightarrow^{64}\rm{Mn}\rightarrow^{64}\rm{Cr}$ and
 $^{64}\rm{Cr}\rightarrow^{64}\rm{V}\rightarrow^{64}\rm{Ti}$
 compositional transitions occur, schematically indicated with
 respect to deep crustal heating~\cite{Stei12} and the carbon
 ignition layer where x-ray superbursts are powered~\cite{Keek12} in the
 right panel, where the neutron star crust
 nuclear reaction network and quasiparticle random phase
 approximation (QRPA) Gamow-Teller
 transition strength distributions reported in Ref.~\cite{Scha14} were used.
 The calculations corresponding to the black and red lines employed
 the 1995 FRDM~\cite{Moll95} and HFB-21~\cite{Gori10} global mass
 models for nuclides with unknown masses, where the 2012 Atomic Mass
 Evaluation~\cite{Audi12} was used otherwise. Calculations indicated
 by solid lines included the mass of $^{64}$Cr presented here.
 \label{fig:IntegratedHeatingCr64}}
 \end{figure*}
 
We have also examined the summed occupancies of the neutron intruder orbitals $g_{9/2}$ and $d_{5/2}$ 
within the LNPS' model. The contribution of those is shown in Fig.
\ref{fig:CrS2nE2plusTrends}, added to the GXPF1A
results. The occupation of the neutron intruder orbitals becomes
significant at $N=36$ ($\sim$1.8 particles) and coincides 
directly with the place where the deviation between GXPF1A calculations and experiment becomes large.
Further increase of this occupancy with increasing neutron number
(see also Tab.~II of Ref.~\cite{Lenz10}) explains the failure of 
shell-model calculations limited to the $fp$-shell model-space
to reproduce $S_{2n}$ for the neutron-rich chromium isotopes.

\subsection{$A=64$ electron capture heating in the accreted neutron star crust}
\label{ssec:CrAstro}

Heating and cooling due to electron capture reactions within the
accreted neutron star crust have been shown to affect the 
outer crust thermal profile and the associated
astronomical observables~\cite{Gupt07,Estr11,Scha14,Meis15b,Deib15,Deib15b}.
Recent calculations with a
state-of-the-art multi-zone x-ray burst model have shown that $A=64$
nuclides
dominate the crust composition for a wide set of astrophysical
conditions (and varied nuclear physics assumptions) that are thought to correspond to typical x-ray bursting
systems~\cite{Cybu16} and previous work has also demonstrated large
$A=64$ production for stable-burning and superbursting
systems~\cite{Scha99,Scha03}.
In large part due to this prevalence, Ref.~\cite{Gupt07}
identified the
$^{64}\rm{Cr}\rightarrow^{64}\rm{V}\rightarrow^{64}\rm{Ti}$
electron-capture sequence as one of the main sources of heat (along
with neutron-capture reactions) at the lower extent 
of the outer crust (i.e. at electron Fermi energy
$E_{\rm{F}}\geq18.5$~MeV). Though weaker than deep crustal
heating sources~\cite{Stei12}, the shallower depth of this heat
source makes it important to consider when calculating the layer at
which carbon ignites to power x-ray superbursts, as shown
schematically in the right panel of
Fig.~\ref{fig:IntegratedHeatingCr64}.

We performed calculations with a crust composition evolution
model~\cite{Gupt07,Scha14}
in order to assess the impact of our newly measured $^{64}$Cr mass
on heat release in the accreted neutron star outer crust.
The model evolves the composition of an accreted fluid element via
nuclear reactions with increasing pressure $p=\dot{m}gt$ (and
therefore increasing $E_{\rm{F}}$), where the accretion rate
$\dot{m}=2.64\times10^{4}$g$\,$cm$^{-2}\,$s$^{-1}$
($\approx$0.3~$\dot{M}_{\rm{Eddington}}$ for a 10~km radius 1.4~solar~mass
neutron star), surface gravity
$g=1.85\times10^{14}$cm$\,$s$^{-2}$, and time $t$, 
at a constant
temperature of $T=0.5$~GK, mimicking the effect of a fluid element
being naturally buried into the crust via subsequently accreted material.
The crust temperature corresponds to the equilibrium value
calculated by Ref.~\cite{Gupt07} and the astrophysical conditions
are within the range inferred for the present population of observed
formerly-accreting cooling neutron stars~\cite{Turl15}. The nuclear
reaction network includes electron-capture, $\beta$-decay, neutron
capture and emission, and fusion reactions.

The resultant integrated nuclear energy release profiles as a function of depth into the
neutron star from our calculations are shown in
Fig.~\ref{fig:IntegratedHeatingCr64} using our $^{64}$Cr mass and
the $^{64}$Cr masses from the commonly used global mass models
FRDM'95~\cite{Moll95} and HFB-21~\cite{Gori10}. The more than 1~MeV
reduction in binding we observe for $^{64}$Cr with respect to FRDM
and HFB-21, a $\sim3\sigma$ deviation using our experimental
uncertainty, results in a substantially reduced odd-even mass
staggering for both the Fe--Mn--Cr and Cr--V--Ti $A=64$ sequences,
which reduces the heat release from both electron capture
sequences~\cite{Gupt07,Meis15b}. Additionally, the reduced $^{64}$Cr
binding energy leads to an earlier transition to $^{64}$Cr and
therefore a shallower depth for the heat deposition from the 
$^{64}\rm{Cr}\rightarrow^{64}\rm{V}\rightarrow^{64}\rm{Ti}$
electron-capture sequence. We note however that the masses of
$^{64}$V and $^{64}$Ti must be experimentally determined to confirm
our conclusions for this second electron-capture sequence.

\section{Conclusions}
\label{sec:Conclusions}

We performed time-of-flight nuclear mass measurements of the
$A=59-64$ isotopes of chromium at the NSCL at Michigan State
University, where the mass of $^{64}$Cr was determined for the first
time.
Our results demonstrate a different behavior
with respect to the 2012 AME for the $S_{2n}$ trend in the chromium
isotopes approaching $N=40$, indicating the shape transition from
spherical to deformed begins at $N=38$ rather than $N=36$.
This $S_{2n}$ trend difference is primarily due to the discrepancy
between our measured and the 2012 AME evaluated masses for
$^{59-61}$Cr. 
Our $^{64}$Cr mass extends the $S_{2n}$ trend for the
chromium isotopes to $N=40$ for the first time, revealing a trend in
mass systematics which is consistent with the previously inferred
collective behavior of chromium in this region. 
We find a reduction in binding energy for $^{64}$Cr of 1.47~MeV and
1.25~MeV with respect to the global mass models FRDM'95 and HFB-21,
respectively, which are commonly used in astrophysics simulations.
Based on our experimental mass uncertainty, these differences correspond to 
a $\sim3\sigma$ deviation.
This reduction in binding energy
leads to a reduced odd-even mass stagger near chromium in the $A=64$
isobaric chain, ultimately causing a reduction of the magnitude and
depth of electron-capture heating associated with $^{64}$Cr, a major
heat source in the outer crust of accreting neutron stars.
Additionally, we performed state-of-the-art shell-model calculations
to calculate $S_{2n}$ for the chromium isotopic chain,
demonstrating the importance of including the
$g_{9/2}$ and $d_{5/2}$ neutron valence spaces for shell-model
calculations in this region.
Future high-precision (e.g.
Penning trap) mass
measurements of scandium and vanadium isotopes in this region will
enable a reevaluation of the presented data, likely reducing the
systematic uncertainty of our chromium masses. In order to
conclusively determine the magnitude of electron-capture heating in
the $A=64$ isobaric chain, the masses of $^{64}$V and $^{64}$Ti will
need to be measured.


\begin{acknowledgments}
We thank C.~Morse,
C.~Prokop, and J.~Vredevoogd for many useful discussions.
This project is funded by the NSF through Grants No. PHY-0822648,
PHY-1102511, PHY-1404442, and No. PHY-1430152.
S.G. acknowledges support from the DFG under Contracts No.
GE2183/1-1 and No. GE2183/2-1. 
\end{acknowledgments}

\bibliography{TOFmassCrDetailedMeisel}

\begin{thebibliography}{78}%
\makeatletter
\providecommand \@ifxundefined [1]{%
 \@ifx{#1\undefined}
}%
\providecommand \@ifnum [1]{%
 \ifnum #1\expandafter \@firstoftwo
 \else \expandafter \@secondoftwo
 \fi
}%
\providecommand \@ifx [1]{%
 \ifx #1\expandafter \@firstoftwo
 \else \expandafter \@secondoftwo
 \fi
}%
\providecommand \natexlab [1]{#1}%
\providecommand \enquote  [1]{``#1''}%
\providecommand \bibnamefont  [1]{#1}%
\providecommand \bibfnamefont [1]{#1}%
\providecommand \citenamefont [1]{#1}%
\providecommand \href@noop [0]{\@secondoftwo}%
\providecommand \href [0]{\begingroup \@sanitize@url \@href}%
\providecommand \@href[1]{\@@startlink{#1}\@@href}%
\providecommand \@@href[1]{\endgroup#1\@@endlink}%
\providecommand \@sanitize@url [0]{\catcode `\\12\catcode `\$12\catcode
  `\&12\catcode `\#12\catcode `\^12\catcode `\_12\catcode `\%12\relax}%
\providecommand \@@startlink[1]{}%
\providecommand \@@endlink[0]{}%
\providecommand \url  [0]{\begingroup\@sanitize@url \@url }%
\providecommand \@url [1]{\endgroup\@href {#1}{\urlprefix }}%
\providecommand \urlprefix  [0]{URL }%
\providecommand \Eprint [0]{\href }%
\providecommand \doibase [0]{http://dx.doi.org/}%
\providecommand \selectlanguage [0]{\@gobble}%
\providecommand \bibinfo  [0]{\@secondoftwo}%
\providecommand \bibfield  [0]{\@secondoftwo}%
\providecommand \translation [1]{[#1]}%
\providecommand \BibitemOpen [0]{}%
\providecommand \bibitemStop [0]{}%
\providecommand \bibitemNoStop [0]{.\EOS\space}%
\providecommand \EOS [0]{\spacefactor3000\relax}%
\providecommand \BibitemShut  [1]{\csname bibitem#1\endcsname}%
\let\auto@bib@innerbib\@empty
\bibitem [{\citenamefont {Brown}(2001)}]{Brow01}%
  \BibitemOpen
  \bibfield  {author} {\bibinfo {author} {\bibfnamefont {B.~A.}\ \bibnamefont
  {Brown}},\ }\href {\doibase http://dx.doi.org/10.1016/S0146-6410(01)00159-4}
  {\bibfield  {journal} {\bibinfo  {journal} {Prog. Part. Nucl. Phys.}\
  }\textbf {\bibinfo {volume} {47}},\ \bibinfo {pages} {517 } (\bibinfo {year}
  {2001})}\BibitemShut {NoStop}%
\bibitem [{\citenamefont {Janssens}(2013)}]{Jans13}%
  \BibitemOpen
  \bibfield  {author} {\bibinfo {author} {\bibfnamefont {R.~V.~F.}\
  \bibnamefont {Janssens}},\ }\href
  {http://stacks.iop.org/1402-4896/2013/i=T152/a=014005} {\bibfield  {journal}
  {\bibinfo  {journal} {Phys. Scr.}\ }\textbf {\bibinfo {volume} {T152}},\
  \bibinfo {pages} {014005} (\bibinfo {year} {2013})}\BibitemShut {NoStop}%
\bibitem [{\citenamefont {Erler}\ \emph {et~al.}(2012)\citenamefont {Erler}
  \emph {et~al.}}]{Erle12}%
  \BibitemOpen
  \bibfield  {author} {\bibinfo {author} {\bibfnamefont {J.}~\bibnamefont
  {Erler}} \emph {et~al.},\ }\href {\doibase
  http://dx.doi.org/10.1038/nature11188} {\bibfield  {journal} {\bibinfo
  {journal} {Nature (London)}\ }\textbf {\bibinfo {volume} {486}},\ \bibinfo
  {pages} {509} (\bibinfo {year} {2012})}\BibitemShut {NoStop}%
\bibitem [{\citenamefont {Sorlin}\ \emph {et~al.}(2003)\citenamefont {Sorlin}
  \emph {et~al.}}]{Sorl03}%
  \BibitemOpen
  \bibfield  {author} {\bibinfo {author} {\bibfnamefont {O.}~\bibnamefont
  {Sorlin}} \emph {et~al.},\ }\href {\doibase
  http://dx.doi.org/10.1140/epja/i2002-10069-9} {\bibfield  {journal} {\bibinfo
   {journal} {Euro. Phys. Jour. A}\ }\textbf {\bibinfo {volume} {16}},\
  \bibinfo {pages} {55} (\bibinfo {year} {2003})}\BibitemShut {NoStop}%
\bibitem [{\citenamefont {Gade}\ \emph {et~al.}(2010)\citenamefont {Gade} \emph
  {et~al.}}]{Gade10}%
  \BibitemOpen
  \bibfield  {author} {\bibinfo {author} {\bibfnamefont {A.}~\bibnamefont
  {Gade}} \emph {et~al.},\ }\href {\doibase 10.1103/PhysRevC.81.051304}
  {\bibfield  {journal} {\bibinfo  {journal} {Phys. Rev. C}\ }\textbf {\bibinfo
  {volume} {81}},\ \bibinfo {pages} {051304(R)} (\bibinfo {year}
  {2010})}\BibitemShut {NoStop}%
\bibitem [{\citenamefont {Baugher}\ \emph {et~al.}(2012)\citenamefont {Baugher}
  \emph {et~al.}}]{Baug12}%
  \BibitemOpen
  \bibfield  {author} {\bibinfo {author} {\bibfnamefont {T.}~\bibnamefont
  {Baugher}} \emph {et~al.},\ }\href {\doibase 10.1103/PhysRevC.86.011305}
  {\bibfield  {journal} {\bibinfo  {journal} {Phys. Rev. C}\ }\textbf {\bibinfo
  {volume} {86}},\ \bibinfo {pages} {011305(R)} (\bibinfo {year}
  {2012})}\BibitemShut {NoStop}%
\bibitem [{\citenamefont {Naimi}\ \emph {et~al.}(2012)\citenamefont {Naimi}
  \emph {et~al.}}]{Naim12}%
  \BibitemOpen
  \bibfield  {author} {\bibinfo {author} {\bibfnamefont {S.}~\bibnamefont
  {Naimi}} \emph {et~al.},\ }\href {\doibase 10.1103/PhysRevC.86.014325}
  {\bibfield  {journal} {\bibinfo  {journal} {Phys. Rev. C}\ }\textbf {\bibinfo
  {volume} {86}},\ \bibinfo {pages} {014325} (\bibinfo {year}
  {2012})}\BibitemShut {NoStop}%
\bibitem [{\citenamefont {Crawford}\ \emph {et~al.}(2013)\citenamefont
  {Crawford} \emph {et~al.}}]{Craw13}%
  \BibitemOpen
  \bibfield  {author} {\bibinfo {author} {\bibfnamefont {H.~L.}\ \bibnamefont
  {Crawford}} \emph {et~al.},\ }\href {\doibase 10.1103/PhysRevLett.110.242701}
  {\bibfield  {journal} {\bibinfo  {journal} {Phys. Rev. Lett.}\ }\textbf
  {\bibinfo {volume} {110}},\ \bibinfo {pages} {242701} (\bibinfo {year}
  {2013})}\BibitemShut {NoStop}%
\bibitem [{\citenamefont {Braunroth}\ \emph {et~al.}(2015)\citenamefont
  {Braunroth} \emph {et~al.}}]{Brau15}%
  \BibitemOpen
  \bibfield  {author} {\bibinfo {author} {\bibfnamefont {T.}~\bibnamefont
  {Braunroth}} \emph {et~al.},\ }\href {\doibase 10.1103/PhysRevC.92.034306}
  {\bibfield  {journal} {\bibinfo  {journal} {Phys. Rev. C}\ }\textbf {\bibinfo
  {volume} {92}},\ \bibinfo {pages} {034306} (\bibinfo {year}
  {2015})}\BibitemShut {NoStop}%
\bibitem [{\citenamefont {Adrich}\ \emph {et~al.}(2008)\citenamefont {Adrich},
  \citenamefont {Amthor}, \citenamefont {Bazin}, \citenamefont {Bowen},
  \citenamefont {Brown}, \citenamefont {Campbell}, \citenamefont {Cook},
  \citenamefont {Gade}, \citenamefont {Galaviz}, \citenamefont {Glasmacher},
  \citenamefont {McDaniel}, \citenamefont {Miller}, \citenamefont {Obertelli},
  \citenamefont {Shimbara}, \citenamefont {Siwek}, \citenamefont {Tostevin},\
  and\ \citenamefont {Weisshaar}}]{Adri08}%
  \BibitemOpen
  \bibfield  {author} {\bibinfo {author} {\bibfnamefont {P.}~\bibnamefont
  {Adrich}}, \bibinfo {author} {\bibfnamefont {A.~M.}\ \bibnamefont {Amthor}},
  \bibinfo {author} {\bibfnamefont {D.}~\bibnamefont {Bazin}}, \bibinfo
  {author} {\bibfnamefont {M.~D.}\ \bibnamefont {Bowen}}, \bibinfo {author}
  {\bibfnamefont {B.~A.}\ \bibnamefont {Brown}}, \bibinfo {author}
  {\bibfnamefont {C.~M.}\ \bibnamefont {Campbell}}, \bibinfo {author}
  {\bibfnamefont {J.~M.}\ \bibnamefont {Cook}}, \bibinfo {author}
  {\bibfnamefont {A.}~\bibnamefont {Gade}}, \bibinfo {author} {\bibfnamefont
  {D.}~\bibnamefont {Galaviz}}, \bibinfo {author} {\bibfnamefont
  {T.}~\bibnamefont {Glasmacher}}, \bibinfo {author} {\bibfnamefont
  {S.}~\bibnamefont {McDaniel}}, \bibinfo {author} {\bibfnamefont
  {D.}~\bibnamefont {Miller}}, \bibinfo {author} {\bibfnamefont
  {A.}~\bibnamefont {Obertelli}}, \bibinfo {author} {\bibfnamefont
  {Y.}~\bibnamefont {Shimbara}}, \bibinfo {author} {\bibfnamefont {K.~P.}\
  \bibnamefont {Siwek}}, \bibinfo {author} {\bibfnamefont {J.~A.}\ \bibnamefont
  {Tostevin}}, \ and\ \bibinfo {author} {\bibfnamefont {D.}~\bibnamefont
  {Weisshaar}},\ }\href {\doibase 10.1103/PhysRevC.77.054306} {\bibfield
  {journal} {\bibinfo  {journal} {Phys. Rev. C}\ }\textbf {\bibinfo {volume}
  {77}},\ \bibinfo {pages} {054306} (\bibinfo {year} {2008})}\BibitemShut
  {NoStop}%
\bibitem [{\citenamefont {Lenzi}\ \emph {et~al.}(2010)\citenamefont {Lenzi},
  \citenamefont {Nowacki}, \citenamefont {Poves},\ and\ \citenamefont
  {Sieja}}]{Lenz10}%
  \BibitemOpen
  \bibfield  {author} {\bibinfo {author} {\bibfnamefont {S.~M.}\ \bibnamefont
  {Lenzi}}, \bibinfo {author} {\bibfnamefont {F.}~\bibnamefont {Nowacki}},
  \bibinfo {author} {\bibfnamefont {A.}~\bibnamefont {Poves}}, \ and\ \bibinfo
  {author} {\bibfnamefont {K.}~\bibnamefont {Sieja}},\ }\href {\doibase
  10.1103/PhysRevC.82.054301} {\bibfield  {journal} {\bibinfo  {journal} {Phys.
  Rev. C}\ }\textbf {\bibinfo {volume} {82}},\ \bibinfo {pages} {054301}
  (\bibinfo {year} {2010})}\BibitemShut {NoStop}%
\bibitem [{\citenamefont {Lunney}\ \emph {et~al.}(2003)\citenamefont {Lunney},
  \citenamefont {Pearson},\ and\ \citenamefont {Thibault}}]{Lunn03}%
  \BibitemOpen
  \bibfield  {author} {\bibinfo {author} {\bibfnamefont {D.}~\bibnamefont
  {Lunney}}, \bibinfo {author} {\bibfnamefont {J.~M.}\ \bibnamefont {Pearson}},
  \ and\ \bibinfo {author} {\bibfnamefont {C.}~\bibnamefont {Thibault}},\
  }\href {\doibase 10.1103/RevModPhys.75.1021} {\bibfield  {journal} {\bibinfo
  {journal} {Rev. Mod. Phys.}\ }\textbf {\bibinfo {volume} {75}},\ \bibinfo
  {pages} {1021} (\bibinfo {year} {2003})}\BibitemShut {NoStop}%
\bibitem [{\citenamefont {Meisel}\ \emph
  {et~al.}(2015{\natexlab{a}})\citenamefont {Meisel} \emph {et~al.}}]{Meis15}%
  \BibitemOpen
  \bibfield  {author} {\bibinfo {author} {\bibfnamefont {Z.}~\bibnamefont
  {Meisel}} \emph {et~al.},\ }\href@noop {} {\bibfield  {journal} {\bibinfo
  {journal} {Phys. Rev. Lett.}\ }\textbf {\bibinfo {volume} {114}},\ \bibinfo
  {pages} {022501} (\bibinfo {year} {2015}{\natexlab{a}})}\BibitemShut
  {NoStop}%
\bibitem [{\citenamefont {Schatz}\ \emph {et~al.}(1999)\citenamefont {Schatz},
  \citenamefont {Bildsten}, \citenamefont {Cumming},\ and\ \citenamefont
  {Wiescher}}]{Scha99}%
  \BibitemOpen
  \bibfield  {author} {\bibinfo {author} {\bibfnamefont {H.}~\bibnamefont
  {Schatz}}, \bibinfo {author} {\bibfnamefont {L.}~\bibnamefont {Bildsten}},
  \bibinfo {author} {\bibfnamefont {A.}~\bibnamefont {Cumming}}, \ and\
  \bibinfo {author} {\bibfnamefont {M.}~\bibnamefont {Wiescher}},\ }\href
  {http://stacks.iop.org/0004-637X/524/i=2/a=1014} {\bibfield  {journal}
  {\bibinfo  {journal} {Astrophys. J.}\ }\textbf {\bibinfo {volume} {524}},\
  \bibinfo {pages} {1014} (\bibinfo {year} {1999})}\BibitemShut {NoStop}%
\bibitem [{\citenamefont {Schatz}\ \emph {et~al.}(2001)\citenamefont {Schatz}
  \emph {et~al.}}]{Scha01}%
  \BibitemOpen
  \bibfield  {author} {\bibinfo {author} {\bibfnamefont {H.}~\bibnamefont
  {Schatz}} \emph {et~al.},\ }\href@noop {} {\bibfield  {journal} {\bibinfo
  {journal} {Phys. Rev. Lett.}\ }\textbf {\bibinfo {volume} {86}},\ \bibinfo
  {pages} {3471} (\bibinfo {year} {2001})}\BibitemShut {NoStop}%
\bibitem [{\citenamefont {Schatz}\ \emph {et~al.}(2003)\citenamefont {Schatz},
  \citenamefont {Bildsten},\ and\ \citenamefont {Cumming}}]{Scha03}%
  \BibitemOpen
  \bibfield  {author} {\bibinfo {author} {\bibfnamefont {H.}~\bibnamefont
  {Schatz}}, \bibinfo {author} {\bibfnamefont {L.}~\bibnamefont {Bildsten}}, \
  and\ \bibinfo {author} {\bibfnamefont {A.}~\bibnamefont {Cumming}},\ }\href
  {\doibase http://dx.doi.org/10.1086/368107} {\bibfield  {journal} {\bibinfo
  {journal} {Astrophys. J. Lett.}\ }\textbf {\bibinfo {volume} {583}},\
  \bibinfo {pages} {L87} (\bibinfo {year} {2003})}\BibitemShut {NoStop}%
\bibitem [{\citenamefont {Gupta}\ \emph {et~al.}(2007)\citenamefont {Gupta},
  \citenamefont {Brown}, \citenamefont {Schatz}, \citenamefont {M\"{o}ller},\
  and\ \citenamefont {Kratz}}]{Gupt07}%
  \BibitemOpen
  \bibfield  {author} {\bibinfo {author} {\bibfnamefont {S.}~\bibnamefont
  {Gupta}}, \bibinfo {author} {\bibfnamefont {E.~F.}\ \bibnamefont {Brown}},
  \bibinfo {author} {\bibfnamefont {H.}~\bibnamefont {Schatz}}, \bibinfo
  {author} {\bibfnamefont {P.}~\bibnamefont {M\"{o}ller}}, \ and\ \bibinfo
  {author} {\bibfnamefont {K.-L.}\ \bibnamefont {Kratz}},\ }\href
  {http://stacks.iop.org/0004-637X/662/i=2/a=1188} {\bibfield  {journal}
  {\bibinfo  {journal} {Astrophys. J.}\ }\textbf {\bibinfo {volume} {662}},\
  \bibinfo {pages} {1188} (\bibinfo {year} {2007})}\BibitemShut {NoStop}%
\bibitem [{\citenamefont {Estrad\'e}\ \emph {et~al.}(2011)\citenamefont
  {Estrad\'e} \emph {et~al.}}]{Estr11}%
  \BibitemOpen
  \bibfield  {author} {\bibinfo {author} {\bibfnamefont {A.}~\bibnamefont
  {Estrad\'e}} \emph {et~al.},\ }\href {\doibase
  10.1103/PhysRevLett.107.172503} {\bibfield  {journal} {\bibinfo  {journal}
  {Phys. Rev. Lett.}\ }\textbf {\bibinfo {volume} {107}},\ \bibinfo {pages}
  {172503} (\bibinfo {year} {2011})}\BibitemShut {NoStop}%
\bibitem [{\citenamefont {Meisel}\ \emph
  {et~al.}(2015{\natexlab{b}})\citenamefont {Meisel} \emph {et~al.}}]{Meis15b}%
  \BibitemOpen
  \bibfield  {author} {\bibinfo {author} {\bibfnamefont {Z.}~\bibnamefont
  {Meisel}} \emph {et~al.},\ }\href@noop {} {\bibfield  {journal} {\bibinfo
  {journal} {Phys. Rev. Lett.}\ }\textbf {\bibinfo {volume} {115}},\ \bibinfo
  {pages} {162501} (\bibinfo {year} {2015}{\natexlab{b}})}\BibitemShut
  {NoStop}%
\bibitem [{\citenamefont {{Woosley}}\ and\ \citenamefont
  {{Taam}}(1976)}]{Woos76}%
  \BibitemOpen
  \bibfield  {author} {\bibinfo {author} {\bibfnamefont {S.~E.}\ \bibnamefont
  {{Woosley}}}\ and\ \bibinfo {author} {\bibfnamefont {R.~E.}\ \bibnamefont
  {{Taam}}},\ }\href {\doibase http://dx.doi.org/10.1038/263101a0} {\bibfield
  {journal} {\bibinfo  {journal} {Nature (London)}\ }\textbf {\bibinfo {volume}
  {263}},\ \bibinfo {pages} {101} (\bibinfo {year} {1976})}\BibitemShut
  {NoStop}%
\bibitem [{\citenamefont {Schatz}\ and\ \citenamefont {Rehm}(2006)}]{Scha06}%
  \BibitemOpen
  \bibfield  {author} {\bibinfo {author} {\bibfnamefont {H.}~\bibnamefont
  {Schatz}}\ and\ \bibinfo {author} {\bibfnamefont {K.}~\bibnamefont {Rehm}},\
  }\href {\doibase http://dx.doi.org/10.1016/j.nuclphysa.2005.05.200}
  {\bibfield  {journal} {\bibinfo  {journal} {Nucl. Phys. A}\ }\textbf
  {\bibinfo {volume} {777}},\ \bibinfo {pages} {601 } (\bibinfo {year}
  {2006})}\BibitemShut {NoStop}%
\bibitem [{\citenamefont {Parikh}\ \emph {et~al.}(2013)\citenamefont {Parikh},
  \citenamefont {Jos\'e}, \citenamefont {Sala},\ and\ \citenamefont
  {Iliadis}}]{Pari13}%
  \BibitemOpen
  \bibfield  {author} {\bibinfo {author} {\bibfnamefont {A.}~\bibnamefont
  {Parikh}}, \bibinfo {author} {\bibfnamefont {J.}~\bibnamefont {Jos\'e}},
  \bibinfo {author} {\bibfnamefont {G.}~\bibnamefont {Sala}}, \ and\ \bibinfo
  {author} {\bibfnamefont {C.}~\bibnamefont {Iliadis}},\ }\href {\doibase
  http://dx.doi.org/10.1016/j.ppnp.2012.11.002} {\bibfield  {journal} {\bibinfo
   {journal} {Prog. Part. Nucl. Phys.}\ }\textbf {\bibinfo {volume} {69}},\
  \bibinfo {pages} {225 } (\bibinfo {year} {2013})}\BibitemShut {NoStop}%
\bibitem [{\citenamefont {Cumming}\ and\ \citenamefont
  {Bildsten}(2001)}]{Cumm01}%
  \BibitemOpen
  \bibfield  {author} {\bibinfo {author} {\bibfnamefont {A.}~\bibnamefont
  {Cumming}}\ and\ \bibinfo {author} {\bibfnamefont {L.}~\bibnamefont
  {Bildsten}},\ }\href {http://stacks.iop.org/1538-4357/559/i=2/a=L127}
  {\bibfield  {journal} {\bibinfo  {journal} {Astrophys. J. Lett.}\ }\textbf
  {\bibinfo {volume} {559}},\ \bibinfo {pages} {L127} (\bibinfo {year}
  {2001})}\BibitemShut {NoStop}%
\bibitem [{\citenamefont {Keek}\ \emph {et~al.}(2012)\citenamefont {Keek},
  \citenamefont {Heger},\ and\ \citenamefont {in~'t Zand}}]{Keek12}%
  \BibitemOpen
  \bibfield  {author} {\bibinfo {author} {\bibfnamefont {L.}~\bibnamefont
  {Keek}}, \bibinfo {author} {\bibfnamefont {A.}~\bibnamefont {Heger}}, \ and\
  \bibinfo {author} {\bibfnamefont {J.~J.~M.}\ \bibnamefont {in~'t Zand}},\
  }\href {http://stacks.iop.org/0004-637X/752/i=2/a=150} {\bibfield  {journal}
  {\bibinfo  {journal} {Astrophys. J.}\ }\textbf {\bibinfo {volume} {752}},\
  \bibinfo {pages} {150} (\bibinfo {year} {2012})}\BibitemShut {NoStop}%
\bibitem [{\citenamefont {Brown}\ and\ \citenamefont {Cumming}(2009)}]{Brow09}%
  \BibitemOpen
  \bibfield  {author} {\bibinfo {author} {\bibfnamefont {E.~F.}\ \bibnamefont
  {Brown}}\ and\ \bibinfo {author} {\bibfnamefont {A.}~\bibnamefont
  {Cumming}},\ }\href {http://stacks.iop.org/0004-637X/698/i=2/a=1020}
  {\bibfield  {journal} {\bibinfo  {journal} {Astrophys. J.}\ }\textbf
  {\bibinfo {volume} {698}},\ \bibinfo {pages} {1020} (\bibinfo {year}
  {2009})}\BibitemShut {NoStop}%
\bibitem [{\citenamefont {Deibel}\ \emph {et~al.}(2015)\citenamefont {Deibel},
  \citenamefont {Cumming}, \citenamefont {Brown},\ and\ \citenamefont
  {Page}}]{Deib15}%
  \BibitemOpen
  \bibfield  {author} {\bibinfo {author} {\bibfnamefont {A.}~\bibnamefont
  {Deibel}}, \bibinfo {author} {\bibfnamefont {A.}~\bibnamefont {Cumming}},
  \bibinfo {author} {\bibfnamefont {E.~F.}\ \bibnamefont {Brown}}, \ and\
  \bibinfo {author} {\bibfnamefont {D.}~\bibnamefont {Page}},\ }\href
  {http://stacks.iop.org/2041-8205/809/i=2/a=L31} {\bibfield  {journal}
  {\bibinfo  {journal} {Astrophys. J. Lett.}\ }\textbf {\bibinfo {volume}
  {809}},\ \bibinfo {pages} {L31} (\bibinfo {year} {2015})}\BibitemShut
  {NoStop}%
\bibitem [{\citenamefont {Bildsten}(1998)}]{Bild98}%
  \BibitemOpen
  \bibfield  {author} {\bibinfo {author} {\bibfnamefont {L.}~\bibnamefont
  {Bildsten}},\ }\href {http://stacks.iop.org/1538-4357/501/i=1/a=L89}
  {\bibfield  {journal} {\bibinfo  {journal} {Astrophys. J. Lett.}\ }\textbf
  {\bibinfo {volume} {501}},\ \bibinfo {pages} {L89} (\bibinfo {year}
  {1998})}\BibitemShut {NoStop}%
\bibitem [{\citenamefont {Ushomirsky}\ \emph {et~al.}(2000)\citenamefont
  {Ushomirsky}, \citenamefont {Cutler},\ and\ \citenamefont
  {Bildsten}}]{Usho00}%
  \BibitemOpen
  \bibfield  {author} {\bibinfo {author} {\bibfnamefont {G.}~\bibnamefont
  {Ushomirsky}}, \bibinfo {author} {\bibfnamefont {C.}~\bibnamefont {Cutler}},
  \ and\ \bibinfo {author} {\bibfnamefont {L.}~\bibnamefont {Bildsten}},\
  }\href {\doibase 10.1046/j.1365-8711.2000.03938.x} {\bibfield  {journal}
  {\bibinfo  {journal} {Mon. Not. R. Astron. Soc.}\ }\textbf {\bibinfo {volume}
  {319}},\ \bibinfo {pages} {902} (\bibinfo {year} {2000})}\BibitemShut
  {NoStop}%
\bibitem [{\citenamefont {Audi}\ \emph {et~al.}(2012)\citenamefont {Audi},
  \citenamefont {Wang}, \citenamefont {Wapstra}, \citenamefont {Kondev},
  \citenamefont {MacCormick}, \citenamefont {Xu},\ and\ \citenamefont
  {Pfeiffer}}]{Audi12}%
  \BibitemOpen
  \bibfield  {author} {\bibinfo {author} {\bibfnamefont {G.}~\bibnamefont
  {Audi}}, \bibinfo {author} {\bibfnamefont {M.}~\bibnamefont {Wang}}, \bibinfo
  {author} {\bibfnamefont {A.}~\bibnamefont {Wapstra}}, \bibinfo {author}
  {\bibfnamefont {F.}~\bibnamefont {Kondev}}, \bibinfo {author} {\bibfnamefont
  {M.}~\bibnamefont {MacCormick}}, \bibinfo {author} {\bibfnamefont
  {X.}~\bibnamefont {Xu}}, \ and\ \bibinfo {author} {\bibfnamefont
  {B.}~\bibnamefont {Pfeiffer}},\ }\href
  {http://stacks.iop.org/1674-1137/36/i=12/a=002} {\bibfield  {journal}
  {\bibinfo  {journal} {Chin. Phys. C}\ }\textbf {\bibinfo {volume} {36}},\
  \bibinfo {pages} {1287} (\bibinfo {year} {2012})}\BibitemShut {NoStop}%
\bibitem [{\citenamefont {Meisel}\ and\ \citenamefont {George}(2013)}]{Meis13}%
  \BibitemOpen
  \bibfield  {author} {\bibinfo {author} {\bibfnamefont {Z.}~\bibnamefont
  {Meisel}}\ and\ \bibinfo {author} {\bibfnamefont {S.}~\bibnamefont
  {George}},\ }\href {\doibase http://dx.doi.org/10.1016/j.ijms.2013.03.022}
  {\bibfield  {journal} {\bibinfo  {journal} {Int. J. Mass Spectrom.}\ }\textbf
  {\bibinfo {volume} {349-350}},\ \bibinfo {pages} {145 } (\bibinfo {year}
  {2013})}\BibitemShut {NoStop}%
\bibitem [{\citenamefont {{Gaudefroy}}\ \emph {et~al.}(2012)\citenamefont
  {{Gaudefroy}} \emph {et~al.}}]{Gaud12}%
  \BibitemOpen
  \bibfield  {author} {\bibinfo {author} {\bibfnamefont {L.}~\bibnamefont
  {{Gaudefroy}}} \emph {et~al.},\ }\href {\doibase
  10.1103/PhysRevLett.109.202503} {\bibfield  {journal} {\bibinfo  {journal}
  {Phys. Rev. Lett.}\ }\textbf {\bibinfo {volume} {109}},\ \bibinfo {pages}
  {202503} (\bibinfo {year} {2012})}\BibitemShut {NoStop}%
\bibitem [{\citenamefont {Mato\v{s}}\ \emph {et~al.}(2012)\citenamefont
  {Mato\v{s}} \emph {et~al.}}]{Mato12}%
  \BibitemOpen
  \bibfield  {author} {\bibinfo {author} {\bibfnamefont {M.}~\bibnamefont
  {Mato\v{s}}} \emph {et~al.},\ }\href {\doibase
  http://dx.doi.org/10.1016/j.nima.2012.08.104} {\bibfield  {journal} {\bibinfo
   {journal} {Nucl. Instrum. Methods Phys. Res., Sect. A}\ }\textbf {\bibinfo
  {volume} {696}},\ \bibinfo {pages} {171 } (\bibinfo {year}
  {2012})}\BibitemShut {NoStop}%
\bibitem [{\citenamefont {{Morrissey}}\ \emph {et~al.}(2003)\citenamefont
  {{Morrissey}}, \citenamefont {{Sherrill}}, \citenamefont {Steiner},
  \citenamefont {Stolz},\ and\ \citenamefont {Wiedenhoever}}]{Morr03}%
  \BibitemOpen
  \bibfield  {author} {\bibinfo {author} {\bibfnamefont {D.~J.}\ \bibnamefont
  {{Morrissey}}}, \bibinfo {author} {\bibfnamefont {B.~M.}\ \bibnamefont
  {{Sherrill}}}, \bibinfo {author} {\bibfnamefont {M.}~\bibnamefont {Steiner}},
  \bibinfo {author} {\bibfnamefont {A.}~\bibnamefont {Stolz}}, \ and\ \bibinfo
  {author} {\bibfnamefont {I.}~\bibnamefont {Wiedenhoever}},\ }\href {\doibase
  http://dx.doi.org/10.1016/S0168-583X(02)01895-5} {\bibfield  {journal}
  {\bibinfo  {journal} {Nucl. Instrum. Methods Phys. Res., Sect. B}\ }\textbf
  {\bibinfo {volume} {204}},\ \bibinfo {pages} {90 } (\bibinfo {year}
  {2003})}\BibitemShut {NoStop}%
\bibitem [{\citenamefont {Bazin}\ \emph {et~al.}(2003)\citenamefont {Bazin},
  \citenamefont {Caggiano}, \citenamefont {Sherrill}, \citenamefont {Yurkon},\
  and\ \citenamefont {Zeller}}]{Bazi03}%
  \BibitemOpen
  \bibfield  {author} {\bibinfo {author} {\bibfnamefont {D.}~\bibnamefont
  {Bazin}}, \bibinfo {author} {\bibfnamefont {J.}~\bibnamefont {Caggiano}},
  \bibinfo {author} {\bibfnamefont {B.}~\bibnamefont {Sherrill}}, \bibinfo
  {author} {\bibfnamefont {J.}~\bibnamefont {Yurkon}}, \ and\ \bibinfo {author}
  {\bibfnamefont {A.}~\bibnamefont {Zeller}},\ }\href {\doibase
  http://dx.doi.org/10.1016/S0168-583X(02)02142-0} {\bibfield  {journal}
  {\bibinfo  {journal} {Nucl. Instrum. Methods Phys. Res., Sect. B}\ }\textbf
  {\bibinfo {volume} {204}},\ \bibinfo {pages} {629 } (\bibinfo {year}
  {2003})}\BibitemShut {NoStop}%
\bibitem [{\citenamefont {Yurkon}\ \emph {et~al.}(1999)\citenamefont {Yurkon}
  \emph {et~al.}}]{Yurk99}%
  \BibitemOpen
  \bibfield  {author} {\bibinfo {author} {\bibfnamefont {J.}~\bibnamefont
  {Yurkon}} \emph {et~al.},\ }\href {\doibase
  http://dx.doi.org/10.1016/S0168-9002(98)00960-7} {\bibfield  {journal}
  {\bibinfo  {journal} {Nucl. Instrum. Methods Phys. Res., Sect. A}\ }\textbf
  {\bibinfo {volume} {422}},\ \bibinfo {pages} {291 } (\bibinfo {year}
  {1999})}\BibitemShut {NoStop}%
\bibitem [{\citenamefont {York}\ \emph {et~al.}(1999)\citenamefont {York} \emph
  {et~al.}}]{York99}%
  \BibitemOpen
  \bibfield  {author} {\bibinfo {author} {\bibfnamefont {R.}~\bibnamefont
  {York}} \emph {et~al.},\ }in\ \href@noop {} {\emph {\bibinfo {booktitle}
  {Cyclotrons and Their Applications 1998}}},\ \bibinfo {editor} {edited by\
  \bibinfo {editor} {\bibfnamefont {E.}~\bibnamefont {Baron}}\ and\ \bibinfo
  {editor} {\bibfnamefont {M.}~\bibnamefont {Liuvin}}}\ (\bibinfo  {publisher}
  {Institute of Physics Publishing},\ \bibinfo {year} {1999})\ pp.\ \bibinfo
  {pages} {687--691},\ \bibinfo {note} {{P}roceedings of the 15th International
  Conference, Caen, France, 14-19 June 1998}\BibitemShut {NoStop}%
\bibitem [{Sai()}]{SaintGobain}%
  \BibitemOpen
  \href@noop {} {\enquote {\bibinfo {title}
  {http://www.crystals.saint-gobain.com},}\ }\bibinfo {note} {Saint-Gobain
  Crystals}\BibitemShut {NoStop}%
\bibitem [{Ham()}]{Hamamatsu}%
  \BibitemOpen
  \href@noop {} {\enquote {\bibinfo {title} {http://www.hamamatsu.com},}\
  }\bibinfo {note} {Hamamatsu Photonics}\BibitemShut {NoStop}%
\bibitem [{Bel()}]{Belden}%
  \BibitemOpen
  \href@noop {} {\enquote {\bibinfo {title} {http://www.belden.com},}\
  }\bibinfo {note} {Belden CDT Inc.}\BibitemShut {Stop}%
\bibitem [{\citenamefont {Meisel}(2015)}]{Meis15phd}%
  \BibitemOpen
  \bibfield  {author} {\bibinfo {author} {\bibfnamefont {Z.}~\bibnamefont
  {Meisel}},\ }\emph {\bibinfo {title} {Extension of the nuclear mass surface
  for neutron-rich isotopes of argon through iron}},\ \href
  {https://publications.nscl.msu.edu/thesis/Meisel2015_381.pdf} {\bibinfo
  {type} {Ph.{D}. thesis}},\ \bibinfo  {school} {Michigan State University,
  East Lansing} (\bibinfo {year} {2015})\BibitemShut {NoStop}%
\bibitem [{\citenamefont {Shapira}\ \emph {et~al.}(2000)\citenamefont
  {Shapira}, \citenamefont {Lewis},\ and\ \citenamefont {Hulett}}]{Shap00}%
  \BibitemOpen
  \bibfield  {author} {\bibinfo {author} {\bibfnamefont {D.}~\bibnamefont
  {Shapira}}, \bibinfo {author} {\bibfnamefont {T.}~\bibnamefont {Lewis}}, \
  and\ \bibinfo {author} {\bibfnamefont {L.}~\bibnamefont {Hulett}},\ }\href
  {\doibase http://dx.doi.org/10.1016/S0168-9002(00)00499-X} {\bibfield
  {journal} {\bibinfo  {journal} {Nucl. Instrum. Methods Phys. Res., Sect. A}\
  }\textbf {\bibinfo {volume} {454}},\ \bibinfo {pages} {409 } (\bibinfo {year}
  {2000})}\BibitemShut {NoStop}%
\bibitem [{Qua()}]{Quantar}%
  \BibitemOpen
  \href@noop {} {\enquote {\bibinfo {title} {http://www.quantar.com},}\
  }\bibinfo {note} {Quantar Technology Inc.}\BibitemShut {Stop}%
\bibitem [{Mag()}]{MagnetSales}%
  \BibitemOpen
  \href@noop {} {\enquote {\bibinfo {title} {http://www.magnetsales.com},}\
  }\bibinfo {note} {Magnet Sales and Manufacturing Inc.}\BibitemShut {Stop}%
\bibitem [{\citenamefont {Jung}\ \emph {et~al.}(1996)\citenamefont {Jung},
  \citenamefont {Rothard}, \citenamefont {Gervais}, \citenamefont {Grandin},
  \citenamefont {Clouvas},\ and\ \citenamefont {W\"unsch}}]{Jung96}%
  \BibitemOpen
  \bibfield  {author} {\bibinfo {author} {\bibfnamefont {M.}~\bibnamefont
  {Jung}}, \bibinfo {author} {\bibfnamefont {H.}~\bibnamefont {Rothard}},
  \bibinfo {author} {\bibfnamefont {B.}~\bibnamefont {Gervais}}, \bibinfo
  {author} {\bibfnamefont {J.-P.}\ \bibnamefont {Grandin}}, \bibinfo {author}
  {\bibfnamefont {A.}~\bibnamefont {Clouvas}}, \ and\ \bibinfo {author}
  {\bibfnamefont {R.}~\bibnamefont {W\"unsch}},\ }\href {\doibase
  10.1103/PhysRevA.54.4153} {\bibfield  {journal} {\bibinfo  {journal} {Phys.
  Rev. A}\ }\textbf {\bibinfo {volume} {54}},\ \bibinfo {pages} {4153}
  (\bibinfo {year} {1996})}\BibitemShut {NoStop}%
\bibitem [{\citenamefont {Landau}\ and\ \citenamefont
  {Lifshitz}(1975)}]{Land75}%
  \BibitemOpen
  \bibfield  {author} {\bibinfo {author} {\bibfnamefont {L.}~\bibnamefont
  {Landau}}\ and\ \bibinfo {author} {\bibfnamefont {E.}~\bibnamefont
  {Lifshitz}},\ }\href@noop {} {\emph {\bibinfo {title} {The Classical Theory
  of Fields}}},\ \bibinfo {edition} {4th}\ ed.\ (\bibinfo  {publisher}
  {Elsevier},\ \bibinfo {year} {1975})\ \bibinfo {note} {{P}art of the
  \emph{Course of Theoretical Physics, Volume 2}}\BibitemShut {NoStop}%
\bibitem [{\citenamefont {{Rogers}}\ \emph {et~al.}(2015)\citenamefont
  {{Rogers}} \emph {et~al.}}]{Roge15}%
  \BibitemOpen
  \bibfield  {author} {\bibinfo {author} {\bibfnamefont {A.~M.}\ \bibnamefont
  {{Rogers}}} \emph {et~al.},\ }\href {\doibase 10.1016/j.nima.2015.05.070}
  {\bibfield  {journal} {\bibinfo  {journal} {Nucl. Instrum. Methods Phys.
  Res., Sect. A}\ }\textbf {\bibinfo {volume} {795}},\ \bibinfo {pages} {325}
  (\bibinfo {year} {2015})}\BibitemShut {NoStop}%
\bibitem [{ROO()}]{ROOT}%
  \BibitemOpen
  \href@noop {} {\enquote {\bibinfo {title} {https://root.cern.ch/drupal/},}\
  }\bibinfo {note} {{ROOT} Data Analysis Framework}\BibitemShut {NoStop}%
\bibitem [{\citenamefont {Wienholtz}\ \emph {et~al.}(2013)\citenamefont
  {Wienholtz} \emph {et~al.}}]{Wien13}%
  \BibitemOpen
  \bibfield  {author} {\bibinfo {author} {\bibfnamefont {F.}~\bibnamefont
  {Wienholtz}} \emph {et~al.},\ }\href {\doibase
  http://dx.doi.org/10.1038/nature12226} {\bibfield  {journal} {\bibinfo
  {journal} {Nature (London)}\ }\textbf {\bibinfo {volume} {498}},\ \bibinfo
  {pages} {346} (\bibinfo {year} {2013})}\BibitemShut {NoStop}%
\bibitem [{NND()}]{NNDC}%
  \BibitemOpen
  \href@noop {} {\enquote {\bibinfo {title} {http://www.nndc.bnl.gov/},}\
  }\bibinfo {note} {National Nuclear Data Center compilation, Accessed January
  2014}\BibitemShut {NoStop}%
\bibitem [{\citenamefont {Lotz}(1970)}]{Lotz70}%
  \BibitemOpen
  \bibfield  {author} {\bibinfo {author} {\bibfnamefont {W.}~\bibnamefont
  {Lotz}},\ }\href {\doibase 10.1364/JOSA.60.000206} {\bibfield  {journal}
  {\bibinfo  {journal} {Journal of the Optical Society of America}\ }\textbf
  {\bibinfo {volume} {60}},\ \bibinfo {pages} {206} (\bibinfo {year}
  {1970})}\BibitemShut {NoStop}%
\bibitem [{\citenamefont {Chen}\ \emph {et~al.}(2012)\citenamefont {Chen} \emph
  {et~al.}}]{Chen12}%
  \BibitemOpen
  \bibfield  {author} {\bibinfo {author} {\bibfnamefont {L.}~\bibnamefont
  {Chen}} \emph {et~al.},\ }\href@noop {} {\bibfield  {journal} {\bibinfo
  {journal} {Nuc. Phys. A}\ }\textbf {\bibinfo {volume} {882}},\ \bibinfo
  {pages} {71} (\bibinfo {year} {2012})}\BibitemShut {NoStop}%
\bibitem [{\citenamefont {Tu}\ \emph {et~al.}(1990)\citenamefont {Tu} \emph
  {et~al.}}]{Tu90}%
  \BibitemOpen
  \bibfield  {author} {\bibinfo {author} {\bibfnamefont {X.}~\bibnamefont {Tu}}
  \emph {et~al.},\ }\href@noop {} {\bibfield  {journal} {\bibinfo  {journal}
  {Z. Phys. A}\ }\textbf {\bibinfo {volume} {337}},\ \bibinfo {pages} {361}
  (\bibinfo {year} {1990})}\BibitemShut {NoStop}%
\bibitem [{\citenamefont {Seifert}\ \emph {et~al.}(1994)\citenamefont {Seifert}
  \emph {et~al.}}]{Seif94}%
  \BibitemOpen
  \bibfield  {author} {\bibinfo {author} {\bibfnamefont {H.}~\bibnamefont
  {Seifert}} \emph {et~al.},\ }\href {\doibase 10.1007/BF01296329} {\bibfield
  {journal} {\bibinfo  {journal} {Z. Phys. A}\ }\textbf {\bibinfo {volume}
  {349}},\ \bibinfo {pages} {25} (\bibinfo {year} {1994})}\BibitemShut
  {NoStop}%
\bibitem [{\citenamefont {Bai}\ \emph {et~al.}(1998)\citenamefont {Bai},
  \citenamefont {Vieira}, \citenamefont {Seifert},\ and\ \citenamefont
  {Wouters}}]{Bai98}%
  \BibitemOpen
  \bibfield  {author} {\bibinfo {author} {\bibfnamefont {Y.}~\bibnamefont
  {Bai}}, \bibinfo {author} {\bibfnamefont {D.~J.}\ \bibnamefont {Vieira}},
  \bibinfo {author} {\bibfnamefont {H.~L.}\ \bibnamefont {Seifert}}, \ and\
  \bibinfo {author} {\bibfnamefont {J.~M.}\ \bibnamefont {Wouters}},\
  }\href@noop {} {\bibfield  {journal} {\bibinfo  {journal} {AIP Conf. Proc.}\
  }\textbf {\bibinfo {volume} {455}} (\bibinfo {year} {1998})}\BibitemShut
  {NoStop}%
\bibitem [{\citenamefont {M\"{o}ller}\ \emph {et~al.}(1995)\citenamefont
  {M\"{o}ller}, \citenamefont {Nix}, \citenamefont {Myers},\ and\ \citenamefont
  {Swiatecki}}]{Moll95}%
  \BibitemOpen
  \bibfield  {author} {\bibinfo {author} {\bibfnamefont {P.}~\bibnamefont
  {M\"{o}ller}}, \bibinfo {author} {\bibfnamefont {J.}~\bibnamefont {Nix}},
  \bibinfo {author} {\bibfnamefont {W.}~\bibnamefont {Myers}}, \ and\ \bibinfo
  {author} {\bibfnamefont {W.}~\bibnamefont {Swiatecki}},\ }\href {\doibase
  http://dx.doi.org/10.1006/adnd.1995.1002} {\bibfield  {journal} {\bibinfo
  {journal} {At. Data Nucl. Data Tables}\ }\textbf {\bibinfo {volume} {59}},\
  \bibinfo {pages} {185 } (\bibinfo {year} {1995})}\BibitemShut {NoStop}%
\bibitem [{\citenamefont {Goriely}\ \emph {et~al.}(2010)\citenamefont
  {Goriely}, \citenamefont {Chamel},\ and\ \citenamefont {Pearson}}]{Gori10}%
  \BibitemOpen
  \bibfield  {author} {\bibinfo {author} {\bibfnamefont {S.}~\bibnamefont
  {Goriely}}, \bibinfo {author} {\bibfnamefont {N.}~\bibnamefont {Chamel}}, \
  and\ \bibinfo {author} {\bibfnamefont {J.~M.}\ \bibnamefont {Pearson}},\
  }\href {\doibase 10.1103/PhysRevC.82.035804} {\bibfield  {journal} {\bibinfo
  {journal} {Phys. Rev. C}\ }\textbf {\bibinfo {volume} {82}},\ \bibinfo
  {pages} {035804} (\bibinfo {year} {2010})}\BibitemShut {NoStop}%
\bibitem [{\citenamefont {Pearson}\ \emph {et~al.}(2011)\citenamefont
  {Pearson}, \citenamefont {Goriely},\ and\ \citenamefont {Chamel}}]{Pear11}%
  \BibitemOpen
  \bibfield  {author} {\bibinfo {author} {\bibfnamefont {J.~M.}\ \bibnamefont
  {Pearson}}, \bibinfo {author} {\bibfnamefont {S.}~\bibnamefont {Goriely}}, \
  and\ \bibinfo {author} {\bibfnamefont {N.}~\bibnamefont {Chamel}},\ }\href
  {\doibase 10.1103/PhysRevC.83.065810} {\bibfield  {journal} {\bibinfo
  {journal} {Phys. Rev. C}\ }\textbf {\bibinfo {volume} {83}},\ \bibinfo
  {pages} {065810} (\bibinfo {year} {2011})}\BibitemShut {NoStop}%
\bibitem [{\citenamefont {Schatz}\ \emph {et~al.}(2014)\citenamefont {Schatz}
  \emph {et~al.}}]{Scha14}%
  \BibitemOpen
  \bibfield  {author} {\bibinfo {author} {\bibfnamefont {H.}~\bibnamefont
  {Schatz}} \emph {et~al.},\ }\href {\doibase
  http://dx.doi.org/10.1038/nature12757} {\bibfield  {journal} {\bibinfo
  {journal} {Nature (London)}\ }\textbf {\bibinfo {volume} {505}},\ \bibinfo
  {pages} {62} (\bibinfo {year} {2014})}\BibitemShut {NoStop}%
\bibitem [{\citenamefont {Honma}\ \emph {et~al.}(2005)\citenamefont {Honma},
  \citenamefont {Otsuka}, \citenamefont {Brown},\ and\ \citenamefont
  {Mizusaki}}]{Honm05}%
  \BibitemOpen
  \bibfield  {author} {\bibinfo {author} {\bibfnamefont {M.}~\bibnamefont
  {Honma}}, \bibinfo {author} {\bibfnamefont {T.}~\bibnamefont {Otsuka}},
  \bibinfo {author} {\bibfnamefont {B.~A.}\ \bibnamefont {Brown}}, \ and\
  \bibinfo {author} {\bibfnamefont {T.}~\bibnamefont {Mizusaki}},\ }\href@noop
  {} {\bibfield  {journal} {\bibinfo  {journal} {Euro. Phys. Jour. A}\ }\textbf
  {\bibinfo {volume} {25}},\ \bibinfo {pages} {499} (\bibinfo {year}
  {2005})}\BibitemShut {NoStop}%
\bibitem [{\citenamefont {Raman}\ \emph {et~al.}(2001)\citenamefont {Raman},
  \citenamefont {\relax{C.W.} \relax{Nestor Jr.}},\ and\ \citenamefont
  {Tikkanen}}]{Rama01}%
  \BibitemOpen
  \bibfield  {author} {\bibinfo {author} {\bibfnamefont {S.}~\bibnamefont
  {Raman}}, \bibinfo {author} {\bibnamefont {\relax{C.W.} \relax{Nestor Jr.}}},
  \ and\ \bibinfo {author} {\bibfnamefont {P.}~\bibnamefont {Tikkanen}},\
  }\href {\doibase http://dx.doi.org/10.1006/adnd.2001.0858} {\bibfield
  {journal} {\bibinfo  {journal} {At. Data Nucl. Data Tables}\ }\textbf
  {\bibinfo {volume} {78}},\ \bibinfo {pages} {1 } (\bibinfo {year}
  {2001})}\BibitemShut {NoStop}%
\bibitem [{\citenamefont {M\u{a}rginean}\ \emph {et~al.}(2006)\citenamefont
  {M\u{a}rginean} \emph {et~al.}}]{Marg06}%
  \BibitemOpen
  \bibfield  {author} {\bibinfo {author} {\bibfnamefont {N.}~\bibnamefont
  {M\u{a}rginean}} \emph {et~al.},\ }\href {\doibase
  http://dx.doi.org/10.1016/j.physletb.2005.12.047} {\bibfield  {journal}
  {\bibinfo  {journal} {Phys. Lett. B}\ }\textbf {\bibinfo {volume} {633}},\
  \bibinfo {pages} {696 } (\bibinfo {year} {2006})}\BibitemShut {NoStop}%
\bibitem [{\citenamefont {Werner}\ \emph {et~al.}(2016)\citenamefont {Werner}
  \emph {et~al.}}]{Wern16}%
  \BibitemOpen
  \bibfield  {author} {\bibinfo {author} {\bibfnamefont {V.}~\bibnamefont
  {Werner}} \emph {et~al.},\ }\href {\doibase
  http://dx.doi.org/10.1051/epjconf/201610703007} {\bibfield  {journal}
  {\bibinfo  {journal} {Euro. Phys. Jour. Web Conf.}\ }\textbf {\bibinfo
  {volume} {107}},\ \bibinfo {pages} {03007} (\bibinfo {year} {2016})},\
  \bibinfo {note} {{P}roceedings of the International Conference on Nuclear
  Structure and Related Topics 2015, Dubna, Russia}\BibitemShut {NoStop}%
\bibitem [{\citenamefont {Rosenbusch}\ \emph {et~al.}(2015)\citenamefont
  {Rosenbusch} \emph {et~al.}}]{Rose15}%
  \BibitemOpen
  \bibfield  {author} {\bibinfo {author} {\bibfnamefont {M.}~\bibnamefont
  {Rosenbusch}} \emph {et~al.},\ }\href {\doibase
  10.1103/PhysRevLett.114.202501} {\bibfield  {journal} {\bibinfo  {journal}
  {Phys. Rev. Lett.}\ }\textbf {\bibinfo {volume} {114}},\ \bibinfo {pages}
  {202501} (\bibinfo {year} {2015})}\BibitemShut {NoStop}%
\bibitem [{\citenamefont {Iachello}\ \emph {et~al.}(2011)\citenamefont
  {Iachello}, \citenamefont {Leviatan},\ and\ \citenamefont
  {Petrellis}}]{Iach11}%
  \BibitemOpen
  \bibfield  {author} {\bibinfo {author} {\bibfnamefont {F.}~\bibnamefont
  {Iachello}}, \bibinfo {author} {\bibfnamefont {A.}~\bibnamefont {Leviatan}},
  \ and\ \bibinfo {author} {\bibfnamefont {D.}~\bibnamefont {Petrellis}},\
  }\href {\doibase 10.1016/j.physletb.2011.10.024} {\bibfield  {journal}
  {\bibinfo  {journal} {Phys. Lett. B}\ }\textbf {\bibinfo {volume} {705}},\
  \bibinfo {pages} {379} (\bibinfo {year} {2011})}\BibitemShut {NoStop}%
\bibitem [{\citenamefont {Chaudhuri}\ \emph {et~al.}(2013)\citenamefont
  {Chaudhuri} \emph {et~al.}}]{Chau13}%
  \BibitemOpen
  \bibfield  {author} {\bibinfo {author} {\bibfnamefont {A.}~\bibnamefont
  {Chaudhuri}} \emph {et~al.},\ }\href {\doibase 10.1103/PhysRevC.88.054317}
  {\bibfield  {journal} {\bibinfo  {journal} {Phys. Rev. C}\ }\textbf {\bibinfo
  {volume} {88}},\ \bibinfo {pages} {054317} (\bibinfo {year}
  {2013})}\BibitemShut {NoStop}%
\bibitem [{\citenamefont {Pritychenko}\ \emph {et~al.}(1999)\citenamefont
  {Pritychenko} \emph {et~al.}}]{Prit99}%
  \BibitemOpen
  \bibfield  {author} {\bibinfo {author} {\bibfnamefont {B.~V.}\ \bibnamefont
  {Pritychenko}} \emph {et~al.},\ }\href {\doibase
  http://dx.doi.org/10.1016/S0370-2693(99)00850-3} {\bibfield  {journal}
  {\bibinfo  {journal} {Phys. Lett. B}\ }\textbf {\bibinfo {volume} {461}},\
  \bibinfo {pages} {322 } (\bibinfo {year} {1999})}\BibitemShut {NoStop}%
\bibitem [{\citenamefont {Terry}\ \emph {et~al.}(2008)\citenamefont {Terry}
  \emph {et~al.}}]{Terr08}%
  \BibitemOpen
  \bibfield  {author} {\bibinfo {author} {\bibfnamefont {J.~R.}\ \bibnamefont
  {Terry}} \emph {et~al.},\ }\href {\doibase 10.1103/PhysRevC.77.014316}
  {\bibfield  {journal} {\bibinfo  {journal} {Phys. Rev. C}\ }\textbf {\bibinfo
  {volume} {77}},\ \bibinfo {pages} {014316} (\bibinfo {year}
  {2008})}\BibitemShut {NoStop}%
\bibitem [{\citenamefont {Wimmer}\ \emph {et~al.}(2010)\citenamefont {Wimmer}
  \emph {et~al.}}]{Wimm10}%
  \BibitemOpen
  \bibfield  {author} {\bibinfo {author} {\bibfnamefont {K.}~\bibnamefont
  {Wimmer}} \emph {et~al.},\ }\href {\doibase 10.1103/PhysRevLett.105.252501}
  {\bibfield  {journal} {\bibinfo  {journal} {Phys. Rev. Lett.}\ }\textbf
  {\bibinfo {volume} {105}},\ \bibinfo {pages} {252501} (\bibinfo {year}
  {2010})}\BibitemShut {NoStop}%
\bibitem [{\citenamefont {Deacon}\ \emph {et~al.}(2005)\citenamefont {Deacon}
  \emph {et~al.}}]{Deac05}%
  \BibitemOpen
  \bibfield  {author} {\bibinfo {author} {\bibfnamefont {A.}~\bibnamefont
  {Deacon}} \emph {et~al.},\ }\href {\doibase
  http://dx.doi.org/10.1016/j.physletb.2005.07.005} {\bibfield  {journal}
  {\bibinfo  {journal} {Phys. Lett. B}\ }\textbf {\bibinfo {volume} {622}},\
  \bibinfo {pages} {151 } (\bibinfo {year} {2005})}\BibitemShut {NoStop}%
\bibitem [{\citenamefont {Zhu}\ \emph {et~al.}(2006)\citenamefont {Zhu} \emph
  {et~al.}}]{Zhu06}%
  \BibitemOpen
  \bibfield  {author} {\bibinfo {author} {\bibfnamefont {S.}~\bibnamefont
  {Zhu}} \emph {et~al.},\ }\href {\doibase 10.1103/PhysRevC.74.064315}
  {\bibfield  {journal} {\bibinfo  {journal} {Phys. Rev. C}\ }\textbf {\bibinfo
  {volume} {74}},\ \bibinfo {pages} {064315} (\bibinfo {year}
  {2006})}\BibitemShut {NoStop}%
\bibitem [{\citenamefont {Togashi}\ \emph {et~al.}(2015)\citenamefont
  {Togashi}, \citenamefont {Shimizu}, \citenamefont {Utsuno}, \citenamefont
  {Otsuka},\ and\ \citenamefont {Honma}}]{Toga15}%
  \BibitemOpen
  \bibfield  {author} {\bibinfo {author} {\bibfnamefont {T.}~\bibnamefont
  {Togashi}}, \bibinfo {author} {\bibfnamefont {N.}~\bibnamefont {Shimizu}},
  \bibinfo {author} {\bibfnamefont {Y.}~\bibnamefont {Utsuno}}, \bibinfo
  {author} {\bibfnamefont {T.}~\bibnamefont {Otsuka}}, \ and\ \bibinfo {author}
  {\bibfnamefont {M.}~\bibnamefont {Honma}},\ }\href {\doibase
  10.1103/PhysRevC.91.024320} {\bibfield  {journal} {\bibinfo  {journal} {Phys.
  Rev. C}\ }\textbf {\bibinfo {volume} {91}},\ \bibinfo {pages} {024320}
  (\bibinfo {year} {2015})}\BibitemShut {NoStop}%
\bibitem [{\citenamefont {Sahin}\ \emph {et~al.}(2015)\citenamefont {Sahin}
  \emph {et~al.}}]{Sahin2015}%
  \BibitemOpen
  \bibfield  {author} {\bibinfo {author} {\bibfnamefont {E.}~\bibnamefont
  {Sahin}} \emph {et~al.},\ }\href@noop {} {\bibfield  {journal} {\bibinfo
  {journal} {Phys. Rev. C}\ }\textbf {\bibinfo {volume} {91}},\ \bibinfo
  {pages} {034302} (\bibinfo {year} {2015})}\BibitemShut {NoStop}%
\bibitem [{\citenamefont {Morfouace}\ \emph {et~al.}(2015)\citenamefont
  {Morfouace} \emph {et~al.}}]{Morfouace15}%
  \BibitemOpen
  \bibfield  {author} {\bibinfo {author} {\bibfnamefont {P.}~\bibnamefont
  {Morfouace}} \emph {et~al.},\ }\href {\doibase
  http://dx.doi.org/10.1016/j.physletb.2015.10.064} {\bibfield  {journal}
  {\bibinfo  {journal} {Phys. Lett. B}\ }\textbf {\bibinfo {volume} {751}},\
  \bibinfo {pages} {306 } (\bibinfo {year} {2015})}\BibitemShut {NoStop}%
\bibitem [{\citenamefont {Suchyta}\ \emph {et~al.}(2014)\citenamefont {Suchyta}
  \emph {et~al.}}]{Such14}%
  \BibitemOpen
  \bibfield  {author} {\bibinfo {author} {\bibfnamefont {S.}~\bibnamefont
  {Suchyta}} \emph {et~al.},\ }\href {\doibase
  http://dx.doi.org/10.1103/PhysRevC.89.034317} {\bibfield  {journal} {\bibinfo
   {journal} {Phys. Rev. C.}\ }\textbf {\bibinfo {volume} {89}},\ \bibinfo
  {pages} {034317} (\bibinfo {year} {2014})}\BibitemShut {NoStop}%
\bibitem [{\citenamefont {Steiner}(2012)}]{Stei12}%
  \BibitemOpen
  \bibfield  {author} {\bibinfo {author} {\bibfnamefont {A.~W.}\ \bibnamefont
  {Steiner}},\ }\href {\doibase 10.1103/PhysRevC.85.055804} {\bibfield
  {journal} {\bibinfo  {journal} {Phys. Rev. C}\ }\textbf {\bibinfo {volume}
  {85}},\ \bibinfo {pages} {055804} (\bibinfo {year} {2012})}\BibitemShut
  {NoStop}%
\bibitem [{\citenamefont {Deibel}\ \emph {et~al.}(2016)\citenamefont {Deibel},
  \citenamefont {Meisel}, \citenamefont {Schatz}, \citenamefont {Brown},\ and\
  \citenamefont {Cumming}}]{Deib15b}%
  \BibitemOpen
  \bibfield  {author} {\bibinfo {author} {\bibfnamefont {A.}~\bibnamefont
  {Deibel}}, \bibinfo {author} {\bibfnamefont {Z.}~\bibnamefont {Meisel}},
  \bibinfo {author} {\bibfnamefont {H.}~\bibnamefont {Schatz}}, \bibinfo
  {author} {\bibfnamefont {E.~F.}\ \bibnamefont {Brown}}, \ and\ \bibinfo
  {author} {\bibfnamefont {A.}~\bibnamefont {Cumming}},\ }\href@noop {}
  {\enquote {\bibinfo {title} {Urca cooling pairs in the neutron star ocean and
  their effect on superburst cooling},}\ } (\bibinfo {year} {2016}),\ \bibinfo
  {note} {\emph{Submitted}.}\BibitemShut {Stop}%
\bibitem [{\citenamefont {Cyburt}\ \emph {et~al.}(2015)\citenamefont {Cyburt},
  \citenamefont {Amthor}, \citenamefont {Heger}, \citenamefont {Johnson},
  \citenamefont {Meisel}, \citenamefont {Schatz},\ and\ \citenamefont
  {Smith}}]{Cybu16}%
  \BibitemOpen
  \bibfield  {author} {\bibinfo {author} {\bibfnamefont {R.~H.}\ \bibnamefont
  {Cyburt}}, \bibinfo {author} {\bibfnamefont {A.~M.}\ \bibnamefont {Amthor}},
  \bibinfo {author} {\bibfnamefont {A.}~\bibnamefont {Heger}}, \bibinfo
  {author} {\bibfnamefont {E.}~\bibnamefont {Johnson}}, \bibinfo {author}
  {\bibfnamefont {Z.}~\bibnamefont {Meisel}}, \bibinfo {author} {\bibfnamefont
  {H.}~\bibnamefont {Schatz}}, \ and\ \bibinfo {author} {\bibfnamefont
  {K.}~\bibnamefont {Smith}},\ }\href@noop {} {\enquote {\bibinfo {title}
  {Dependence of x-ray burst models on nuclear reaction rates},}\ } (\bibinfo
  {year} {2015}),\ \bibinfo {note} {\emph{Submitted}}\BibitemShut {NoStop}%
\bibitem [{\citenamefont {Turlione}\ \emph {et~al.}(2015)\citenamefont
  {Turlione}, \citenamefont {Aguilera},\ and\ \citenamefont {Pons}}]{Turl15}%
  \BibitemOpen
  \bibfield  {author} {\bibinfo {author} {\bibfnamefont {A.}~\bibnamefont
  {Turlione}}, \bibinfo {author} {\bibfnamefont {D.~N.}\ \bibnamefont
  {Aguilera}}, \ and\ \bibinfo {author} {\bibfnamefont {J.~A.}\ \bibnamefont
  {Pons}},\ }\href@noop {} {\bibfield  {journal} {\bibinfo  {journal} {Astron.
  \& Astrophys.}\ }\textbf {\bibinfo {volume} {577}},\ \bibinfo {pages} {A5}
  (\bibinfo {year} {2015})}\BibitemShut {NoStop}%
\end{thebibliography}%

\end{document}